\documentclass[aps,prb,twocolumn,letterpaper,superscriptaddress,floatfix,showpacs]{revtex4-1}

\usepackage{graphicx,amsmath,amsfonts,amssymb,color,ulem,bm,tabularx}
\usepackage{float}
\def\be{\begin{equation}}
\def\ee{\ee{equation}}

\def\baco{Ba$_3$CoSb$_2$O$_9$}

\def\bam{Ba$_3M$Sb$_2$O$_9$}

\def\ba{$^{135,137}$Ba}
\def\baF{$^{135}$Ba}
\def\baS{$^{137}$Ba}

\def\be{\begin{equation}}
\def\ee{\end{equation}}

\def\t1{$T_1^{-1}$}

\def\ita{\textit{a}}
\def\itb{\textit{b}}
\def\itc{\textit{c}}

\usepackage[usenames,dvipsnames,svgnames,table]{xcolor}

\DeclareFontFamily{OT1}{pzc}{}
\DeclareFontShape{OT1}{pzc}{m}{it}{<-> s * [1.2] pzcmi7t}{}
\DeclareMathAlphabet{\mathpzc}{OT1}{pzc}{m}{it}
\newcommand{\ham}{\mathpzc{H}}

\begin{document}

\title{Quantum phase diagram of the $S=1/2$ triangular-lattice antiferromagnet Ba$_3$CoSb$_2$O$_9$}

\author{G. Koutroulakis}
\email{gkoutrou@physics.ucla.edu}
\altaffiliation{current address: Department of Physics $\&$ Astronomy, UCLA, Los Angeles, CA 90095, USA}
\affiliation{Condensed Matter and Magnet Science, MPA-CMMS, Los Alamos National Laboratory, Los Alamos, NM 87545, USA}

\author{T. Zhou}
\affiliation{Department of Physics $\&$ Astronomy, UCLA, Los Angeles, CA 90095, USA}

\author{Y. Kamiya}
\affiliation{iTHES Research Group and Condensed Matter Theory Laboratory, RIKEN, Wako, Saitama 351-0198, Japan}
\affiliation{Theoretical Division, T-4 and CNLS, Los Alamos National Laboratory, Los Alamos, NM 87545, USA}

\author{J. D. Thompson}
\affiliation{Condensed Matter and Magnet Science, MPA-CMMS, Los Alamos National Laboratory, Los Alamos, NM 87545, USA}

\author{H. D. Zhou}
\affiliation{Department of Physics $\&$ Astronomy, University of Tennessee, Knoxville, TN 37996, USA}

\author{C. D. Batista}
\affiliation{Theoretical Division, T-4 and CNLS, Los Alamos National Laboratory, Los Alamos, NM 87545, USA}

\author{S. E. Brown}
\affiliation{Department of Physics $\&$ Astronomy, UCLA, Los Angeles, CA 90095, USA}

\voffset=0.5cm
\begin{abstract}
The magnetic phases of the ideal spin-1/2 triangular-lattice antiferromagnet \baco\ are identified and studied using \ba\ nuclear magnetic resonance (NMR) spectroscopy in magnetic fields ranging to 30T, oriented parallel and near perpendicular to the crystallographic $ab$-plane. For both directions, the saturation field is approximately 33T. Notably, the NMR spectra provide microscopic evidence for the stabilization of an {\it up-up-down} spin configuration for in-plane fields, giving rise to an one-third magnetization plateau ($M_\text{sat}/3$), as well as for a higher field phase transition near to $\sim (3/5)M_\text{sat}$ for both field orientations. Phase transitions are signaled by the evolution of the NMR spectra, and in some cases through spin-lattice relaxation measurements. The results are compared with expectations obtained from a semi-classical energy density modeling, in which quantum effects are incorporated by effective interactions extracted from the spin-wave analysis of the two-dimensional model. The interlayer coupling also plays a significant role in the outcome. Good agreement between the model and the experimental results is achieved, except for the case of fields approaching the saturation value applied along the {\it c}-axis.


\end{abstract}

\maketitle
\section{Introduction}
The study of frustrated quantum magnets has been a central problem in condensed matter physics over the past decades.\cite{Anderson73,Balents10,Frustration11}  Quantum fluctuations lift accidental degeneracies of the classical ($S \to \infty$) limit, resulting in a breadth of exotic ground states and associated excitations. A characteristic example of a geometrically frustrated quantum system is the spin-1/2 triangular lattice Heisenberg antiferromagnet (TLHAF), which is known to order in a 120$^\circ$ state in the absence of a magnetic field.\cite{Huse88,Jolicoeur89} A hallmark of the TLHAF's quantum character is the prediction for the stabilization of an {\it up-up-down} (UUD) spin configuration at $T=0$ for a finite field range,\cite{Nishimori86,Chubukov91} which  corresponds to a  magnetization plateau at one-third of the saturation value $M_\text{sat}$.

Experimentally, there is a dearth of results on the magnetically ordered states of TLHAFs, mainly due to the difficulty of growing regular triangular-lattice materials. In fact, the only spin-1/2 system for which a $M_\text{sat}/3$ plateau has been observed is Cs$_2$CuBr$_4$, in which the triangular lattice is distorted, and the Dzyaloshinskii-Moriya interaction is key for the details of the magnetization process.\cite{Ono03,Fortune09}
Owing to the relatively weak nearest-neighbor exchange interaction, $J/k_B=19.5$K~\cite{Susuki:2013},
and an effective spin $S=1/2$ ionic ground state, \baco\ provides a unique opportunity to study the full range of ordered phases in the TLHAF, because magnetic field strengths of approximately 30T are sufficient to saturate the system.\cite{Shirata:2012,Zhou:2012} Weak interlayer couplings make for long-range order at finite temperature.\cite{Doi:2004}

\baco\ is one of a family of transition metal oxides with the empirical formula \bam, with $M=$ Co, Cu, Ni, Mn, etc. In these compounds, the transition metal ions carry a magnetic moment. For example, the effective spin state is $S=1/2$ for Co, Cu,\cite{Zhou:2011,Nakatsuji:2012es,Quilliam:2012} and $S=1$ for Ni.\cite{Cheng:2011} The crystal structures are similar though not identical, but nevertheless magnetic frustration is common to all. \baco\ crystallizes in a hexagonal structure with space group P6$_3$/{\it mmc}. The Co$^{2+}$ ions are located at (0,0,0), (0,0,1/2), forming layers of regular triangular lattices along the $ab$-plane.\cite{Appendix1} The magnetic properties are associated with a well-separated Kramers doublet ground state.\cite{Shirata:2012,Lines1971,Palii2003}
Electron spin resonance (ESR) measurements reported nearly isotropic $g$-factors $g_\parallel=3.87$ and $g_\perp=3.84$.\cite{Susuki:2013}
The same measurements infer weakly easy-plane exchange interactions.\cite{Susuki:2013}
The 120$^\circ$ ordering occurs at $T_N=3.8$K in zero field.\cite{Doi:2004}

Other field-induced transitions have been observed in fields smaller than the saturation field, and quantum fluctuations are believed to be playing a significant role in determining the details of the associated phases. For instance, the $M_\text{sat}/3$ magnetization plateau is observed for fields 10T\,$<B<$\,15T, associated with a UUD spin structure.\cite{Shirata:2012} Magnetization measurements on a single crystal sample initially reported the emergence of the UUD phase for ${\mathbf B}\parallel\hat{c}$, thus suggesting easy-axis anisotropy.\cite{Zhou:2012} However, similar subsequent experiments, corroborated by ESR data, detected a transition to the UUD phase only  for in-plane fields, casting doubt on the easy-axis interpretation and advocating weakly easy-plane exchange.\cite{Susuki:2013}

It has thus become clear that \baco\ can serve as an excellent, prototypical spin-1/2 TLHAF in contrast to the quantum disordered candidate systems with Cu, Ni magnetic ions.
Nevertheless, taking into account the interlayer coupling is necessary for a complete description. Nuclear magnetic resonance (NMR) spectroscopy poses as an ideal technique for the investigation of this system, since it can provide a local map of the magnetically ordered states.

Here, we describe results from \ba\ NMR measurements, carried out to determine the detailed spin configuration and phases at field $B\leq30$T, which was oriented in the $ab$-plane ($\theta=\pi/2$) and along the $\hat{c}$-axis ($\theta=0$). Our results for the phase transition lines in the magnetic field-temperature plane agree quantitatively with previous magnetization and specific heat experiments.\cite{Zhou:2012,Susuki:2013} In addition, the NMR probe provides direct microscopic evidence for the transition to the plateau UUD phase for ${\mathbf B}\perp\hat{c}$ in the field range 10-15T. The lower field phase is identified as coplanar to the $ab$-plane, with spin configurations in adjacent layers characterized by a reflection symmetry about the field direction. At greater fields, and beyond the UUD phase, an additional transition between two distinctive coplanar phases is detected before saturation, corresponding to the point where the magnetization is $\sim (3/5)M_\text{sat}$ . In fields aligned with the $\hat{c}$-axis, the NMR spectra are consistent with an ``umbrella'' phase for low fields. Two phase transitions are found with increasing field prior to saturation, including one at $\sim (3/5)M_\text{sat}$, while the collinear UUD state is absent. 

For the most part, the experimental results are found to be in excellent agreement with the predictions of a semi-classical treatment that we developed for computing the quantum phase diagram of \baco. This treatment is inspired by the work of Griset {\it et al.}~\cite{Deformed2011Griset} and it incorporates the effect of quantum fluctuations via the generation of effective coupling constants for the classical spins. These constants are computed by expanding the energy of different spin configurations, which are degenerate solutions in the classical limit in $d=2$, to leading order in $1/S$. We begin by describing this approach and present the consequences, thus providing some context for the results of the NMR measurements.

\section{Theory}

Spin-orbit coupling splits the $^4$T$_1$ ground state of the Co$^{2+}$ ion in a perfect octahedral ligand field into the doublet ${\cal J}=1/2$ (an irreducible representation $\Gamma_6$), quartet ${\cal J} = 3/2$ ($\Gamma_8$), and sextet $\cal J = 5/2$ ($\Gamma_7 + \Gamma_8$) states (${\cal J}$ is the total angular momentum). The lowest energy Kramers doublet ${\cal J}=1/2$ ($\Gamma_6$), is separated from the ${\cal J} = 3/2$ ($\Gamma_8$) quartet by a gap of the order of 200--300K.~\cite{Palii2003} Because this gap is much larger than the exchange interaction between different Co$^{2+}$ moments, it is possible to derive an effective $S=1/2$ spin Hamiltonian for the low-energy doublets,\cite{Lines1971,Palii2003}
which can be decomposed in the following way:
\begin{align}
  \ham = \sum_n \ham_\text{\,2D}^{(n)} + \ham_\text{\,anis} + \ham_\text{\,3D},
\end{align}
where
\begin{align}
  \ham_\text{\,2D}^{(n)} = J\sum_{\langle{ij}\rangle}\mathbf{S}_{n,i} \cdot \mathbf{S}_{n,j} -  g \mu_B \mathbf{B} \cdot \sum_i \mathbf{S}_{n,i}
\end{align}
is the Hamiltonian for a 2D isotropic TLHAF in a magnetic field on the $n$-th layer and {$\langle{ij}\rangle$} runs over the in-plane nearest neighbors. $\ham_\text{\,anis} = (J_\parallel - J) \sum_{n,\langle{ij}\rangle} S_{n,i}^z S_{n,j}^z$ and $\ham_\text{\,3D}$ are small perturbations to the 2D isotropic model, namely the exchange anisotropy and the antiferromagnetic inter-layer coupling, respectively. $J$ ($J_\parallel$) is the in-plane transverse (longitudinal) coupling. According to the ESR measurements, the anisotropy is of the easy-plane type, $J_\parallel / J \approx 0.95$, and the inter-layer exchange is $J' / J \approx 0.025$.\cite{Susuki:2013} We will consider varying the magnetic field direction between $\hat{c}$ ($\hat{z}$ in the spin space) and $\hat{a}$ ($\hat{x}$). For the moment we will assume an isotropic $g$-tensor as suggested by the electron paramagnetic resonance (EPR) measurements.~\cite{Susuki:2013} The effect of $g$-tensor anisotropy will be discussed at the end of the section.

Because the magnitudes of $\ham_\text{\,anis}$ and $\ham_\text{\,3D}$ are rather small relative to $\ham_\text{\,2D}$,  quantum effects  can be included by considering the 2D limit described by $\ham_\text{\,2D}$ (herewith, we omit the layer index $n$ when we refer to an arbitrary layer). The magnetic phase diagram of the quasi-2D system is then determined from a balance between the 2D zero-point energy $\Delta E \propto JS$ and the combined effects of inter-layer coupling $\propto J'S^2$ and anisotropy $\propto(J - J_\parallel)S^2$.
To account for this competition, we first derive the effective classical interaction that is induced by quantum fluctuations. For this purpose, we compute $\Delta E$ for different  classical ground states of $\ham_\text{\,2D}$ to leading order in a $1/S$ expansion. Our procedure is an extension  of a similar approach that was adopted  in Ref.~\onlinecite{Deformed2011Griset}.

\subsection{Effective interaction describing the zero-point energy in $\ham_\text{\,2D}$}
\begin{figure}[t]
  \includegraphics[width=0.8\hsize]{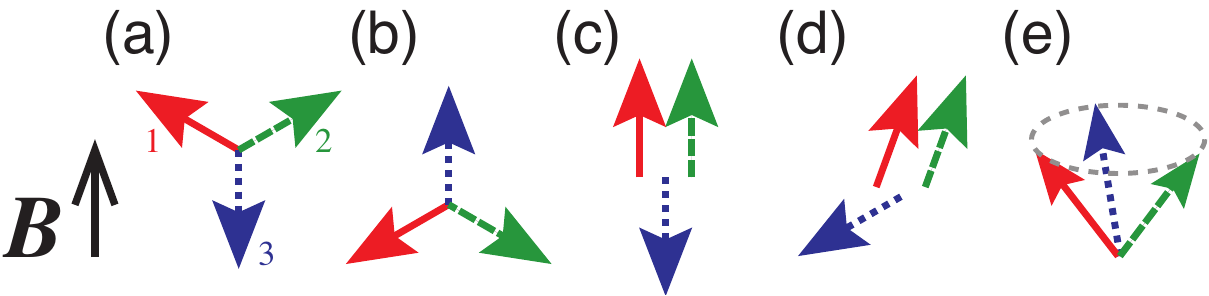}
  \caption{%
    \label{fig:states-2D}
    Representative states for the isotropic TLHAF in 2D: (a) Y, (b) inverted Y, (c) up-up-down, (d) V, and (e) umbrella states.}
\end{figure}

The classical ground state of $\ham_\text{\,2D}$ has an accidental degeneracy in a magnetic field.\cite{Kawamura1985Phase}
We divide the lattice into three sublattices and denote unit vectors of the classical sublattice magnetization as $\bm{\Omega}_{1 \le \mu \le 3}$. The condition for the classical ground state is
\begin{align}
  \bm{\Omega}_1 + \bm{\Omega}_2 + \bm{\Omega}_3 = \frac{g\mu_B \mathbf{B}}{3JS}.
  \label{eq:constraint}
\end{align}
Figure~\ref{fig:states-2D} shows representative classical ground states, namely ``Y,'' ``inverted Y,'' UUD, ``V,'' and umbrella states.
Quantum fluctuations select a particular $T=0$ ordering out of this manifold. It is well-known that the Y/UUD/V states are realized in the low, intermediate, and high field regimes, respectively.\cite{Chubukov91} The $M_\text{sat}/3$ plateau is a manifestation of the UUD state.\cite{Chubukov91}

The zero-point energy, $\Delta E$, of a  given classical ground state $(\bm{\Omega}_1, \bm{\Omega}_2, \bm{\Omega}_3)$  can be computed to lowest order in $1/S$  by expanding the Hamiltonian up to quadratic order in  Holstein-Primakoff bosons (linear spin waves) and calculating the Bogoliubov dispersion $\omega_{\mathbf{k},\nu}$. The index $1 \le \nu \le 3$ denotes the three branches associated with a three-sublattice ordering.
The result is
\begin{align}
  \Delta E (\{\bm{\Omega}_\mu\}) = \frac{1}{2}\sum_{\mathbf{k}\in\text{BZ}} \sum_{1 \le \nu \le 3}\omega_{\mathbf{k},\nu}(\{\bm{\Omega}_\mu\}) - \frac{3N_\text{2D}}{2}JS,
\end{align}
where $N_\text{2D}$ is the total number of sites per layer and BZ stands for the Brillouin zone in 2D.

\begin{figure}
  \includegraphics[width=0.9\hsize]{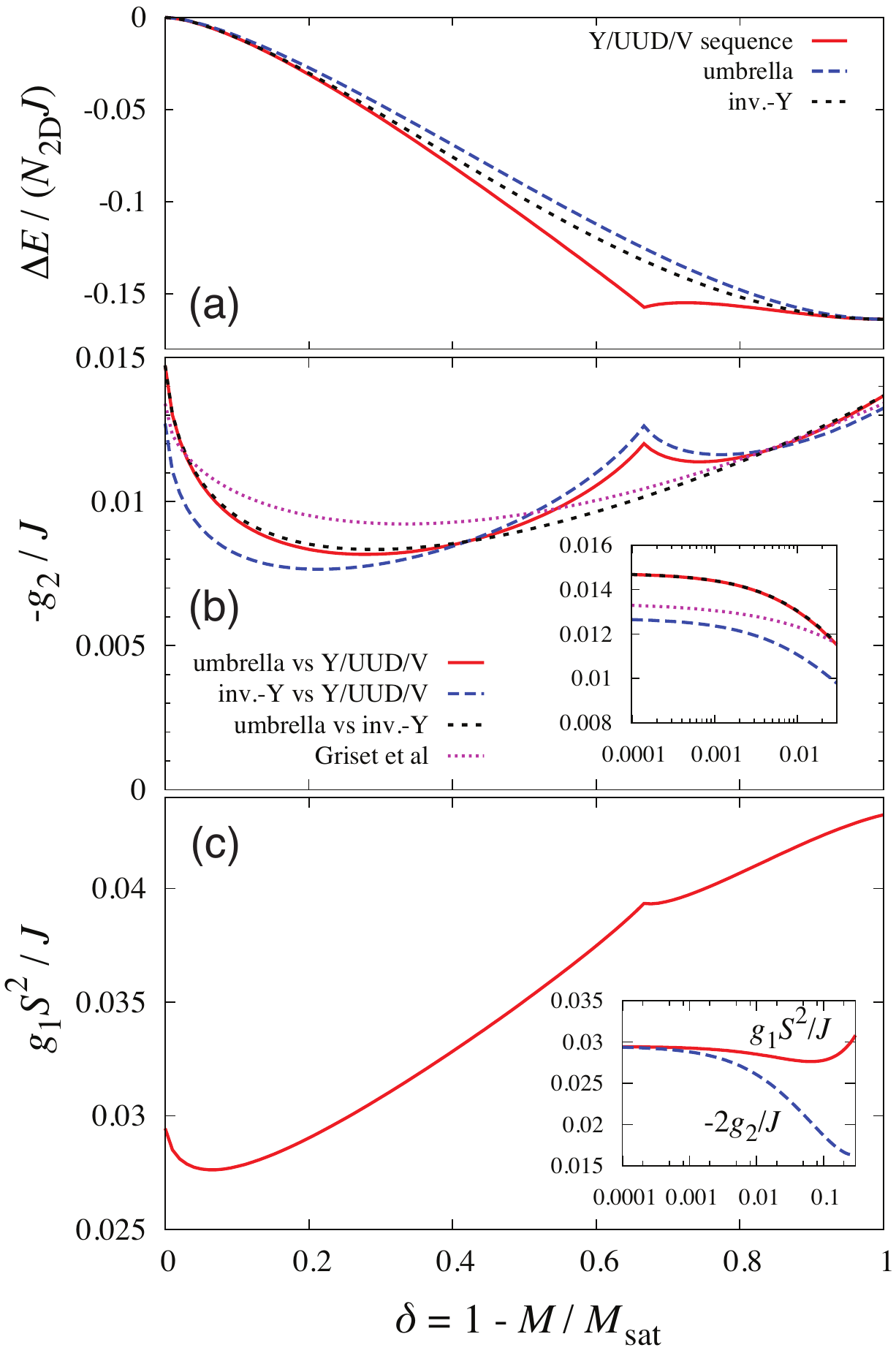}
  \caption{%
    \label{fig:LSW}
Field-dependence of (a) the 2D zero-point energy, (b) the effective biquadratic coupling (with the inset showing a semilogarithmic plot for small $\delta$) obtained by using different combinations of reference states, and (c) the effective correction to the bilinear coupling ($S=1/2$). For comparison, a functional form proposed by Griset,~\textit{et al}.~(Ref.~\onlinecite{Deformed2011Griset}) is also shown in (b). The inset in (c)  confirms the cancellation described by Eq.~\eqref{eq:cancellation}.
  }
\end{figure}

One plausible semiclassical approach for the quasi-2D model is to take $\Delta E$ and add $\ham_\text{\,anis}$ and $\ham_\text{\,3D}$ at the mean-field level, and then perform its minimization in the classical ground state manifold of $\ham_\text{\,2D}$ (i.e., by imposing the constraint Eq.~\eqref{eq:constraint} on each layer). Although this approach is not unrealistic, it is still computationally  expensive to explore a whole parameter space and typically only selected states in the manifold are examined.~\cite{Gekht:1997}
In addition, one has to compute the $1/S$ correction to the spin structure $\bm{\Omega}_\mu \to \widetilde{\bm{\Omega}}_\mu$.
Unlike $\bm{\Omega}_\mu$, $\widetilde{\bm{\Omega}}_\mu$ does not necessarily satisfy the constraint~\eqref{eq:constraint}.

Instead, we propose the following simpler approach, noting that, in principle, $\Delta E (\{\bm{\Omega}_\mu\})$ can be expressed as a polynomial function of $\bm{\Omega}_1$, $\bm{\Omega}_2$, and $\bm{\Omega}_3$, in a form which must have the same symmetry as $\ham_\text{\,2D}$. The different terms of the polynomial expansion can be interpreted as effective interactions for spins on the three sublattices. The computational cost can be greatly reduced by approximating this polynomial form based on a few guiding principles.\cite{Deformed2011Griset} Specifically, we propose
\begin{align}
  \frac{\Delta E(\{\bm{\Omega}_\mu\})}{N_\text{2D}}
  &\approx g_1(M) S^2 \sum_{1 \le \mu \le 3}
  \left(
  \bm{\Omega}_{\mu} \cdot \bm{\Omega}_{\mu+1} - 1
  \right)
  \notag\\
  &\hspace{10pt}+ g_2(M) \sum_{1 \le \mu \le 3}
  \left[
    \left(\bm{\Omega}_{\mu} \cdot \bm{\Omega}_{\mu+1}\right)^2 - 1
  \right]
  \notag\\
  &\equiv
  g_1(M) S^2 F_1(\{\bm{\Omega}_\mu\}) + g_2(M) F_2(\{\bm{\Omega}_\mu\}),
  \label{eq:ansatz}
\end{align}
where $g_1(M)S^2$ and $g_2(M)$ are the effective coupling constants of $O(S)$ and $M=g\mu_B B / (9JS)$ refers to the magnetization in the classical limit. $F_1(\{\bm{\Omega}_\mu\})$ and $F_2(\{\bm{\Omega}_\mu\})$ are defined as $F_1 = F_2 = 0$ for the fully polarized (FP) state, which is an eigenstate of $\ham_\text{\,2D}$ and thus there is no quantum correction. Several previous studies suggest that the negative biquadratic coupling can mimic the quantum effect, as the latter normally favors a collinear spin configuration.\cite{Deformed2011Griset,Chubukov91,Yildirim98}

Within the classical ground state manifold satisfying~\eqref{eq:constraint}, $F_1 = (9/2)[(M/M_\text{sat})^2 - 1]$ is  independent of $\bm{\Omega}_\mu$. Thus, $g_2(M)$ can be extracted from the energy difference between two classical solutions $X$ and $X'$, which we call reference states~\cite{Deformed2011Griset}
\begin{align}
  g_2(M) \approx \frac{\Delta E(\{\bm{\Omega}^{(X)}_{\mu}\}) - \Delta E(\{\bm{\Omega}^{(X')}_{\mu}\})}{N_\text{2D}\left(F_2(\{\bm{\Omega}^{(X)}_{\mu}\}) - F_2(\{\bm{\Omega}^{(X')}_{\mu}\})\right)}.
  \label{eq:g2}
\end{align}
The corresponding results for $X$ and $X'$ being one of (i) Y/UUD/V sequence, (ii) inverted Y, and (iii) umbrella states are shown in Fig.~\ref{fig:LSW} (in producing numerical outputs, we simply extrapolate our results to $S = 1/2$). Except for the cusp at $M = M_\text{sat}/3$, which can be associated with the UUD state, the different choices of reference states give consistent estimates of the same order ($\approx -0.01J$). This observation suggests that Eq.~\eqref{eq:ansatz} is indeed a very good approximation. Furthermore, the negative value of $g_2(M)$ confirms the expectation that the ferro-biquadratic coupling mimics the effect of quantum fluctuations.~\cite{Deformed2011Griset,Chubukov91,Yildirim98} Once $g_2$ is obtained, $g_1$ can be estimated from Eq.~\eqref{eq:ansatz} [see Fig.~\ref{fig:LSW}(c)], where we use $\Delta E$ for the Y/UUD/V sequence and $g_2$ estimated by using the Y/UUD/V and umbrella states as reference states.

Although the asymptotic behavior of $\Delta E \sim \delta^2 (\ln \delta + c_0 + c_1 \delta + \dots)$ and $F_2 \sim \delta$ as $M \to M_{\text{sat}}$ are rather different (here $\delta \equiv (M_\text{sat} - M)/M_\text{sat}$ and $c_0, c_1, \dots$ are constants),\footnote{A logarithmic correction to $\Delta E$ similar to this one was studied in the square lattice Heisenberg model in Ref.~\onlinecite{Zhitomirsky1998Magnetization}.} a cancellation occurs both in the denominator and numerator of Eq.~\eqref{eq:g2}, leading to a consistent asymptotic behavior $\lim_{M \to M_\text{sat}} g_2(M) = \text{const.}$, regardless of the referred classical states [see the inset of Fig.~\ref{fig:LSW}(b)]. For instance, both the denominator and numerator in Eq.~\eqref{eq:g2} scale as $\sim \delta^2$ when the umbrella and V states are used as reference states, while both scale as $\sim \delta^3$ in the case of the inverted Y and V states. On a different note, because $g_2$ remains negative and finite as $M \to M_\text{sat}$,  the ground state selection  by quantum fluctuations survives all the way up to $M = M_\text{sat}$ even though $\Delta E \to 0$ as $M \to M_\text{sat}$. This observation implies that quantum fluctuations 
compete with  other selection mechanisms, such as the inter-layer coupling and anisotropy, even in the high field region $M \lesssim M_\text{sat}$.

Because $F_1$ is independent of $\{\bm{\Omega}_\mu\}$, $g_1$ is not involved in the ground state selection. A main role of $g_1$ is to compensate for the shift of $B_\text{sat}$ produced by $g_2$ (note that the value of $B_\text{sat}$ for $\mathbf{B} \parallel \hat{c}$ is not modified by the presence of quantum fluctuations).
The cancellation of this  shift  requires
\begin{align}
  g_1(M) S^2 + 2 g_2(M) = 0,~~M \to M_\text{sat}.
  \label{eq:cancellation}
\end{align}
Interestingly, as demonstrated in the inset of Fig.~\ref{fig:LSW}(c), this asymptotic behavior automatically follows from Eq.~\eqref{eq:ansatz}: $F_2 \sim 2 F_1 \sim \delta$ and $\Delta E \sim \delta^2 \ln \delta$ imply $g_1 S^2 + 2 g_2 \sim \delta \ln \delta \to 0$ as $\delta \to 0$.

\subsection{Semiclassical approximation for the quasi-2D system}

Based on the above analysis, we propose a six-sublattice expression for the energy density in 3D.   We incorporate the quantum zero-point energy as $M$-dependent bilinear and ferro-biquadratic effective couplings (irrelevant $M$-dependent constants are omitted) and treat $\ham_\text{\,anis}$ and $\ham_\text{\,3D}$ in the mean-field (or classical) approximation:
\begin{align}\label{eq:MF}
  \frac{E_\text{MF}}{N_\text{tot}}
  &= \frac{1}{2} \sum_{\ell=e, o} \sum_{1 \le \mu \le 3}
  \bigl[
    \left(J + g_1(M) \right)S^2
    \bm{\Omega}_{\ell,\mu} \cdot \bm{\Omega}_{\ell,\mu+1}
    \notag\\
    &
    + (J_\parallel - J)S^2 \Omega_{\ell,\mu}^z \Omega_{\ell,\mu+1}^z
    + g_2(M) \left(\bm{\Omega}_{\ell,\mu} \cdot \bm{\Omega}_{\ell,\mu+1}\right)^2
    \bigr]
  \notag\\
  &+ \frac{J' S^2}{3} \sum_{1 \le \mu \le 3} \bm{\Omega}_{e,\mu} \cdot \bm{\Omega}_{o,\mu+1}
  - \frac{g\mu_B S \mathbf{B}}{6} \cdot \sum_{\ell, \mu} \bm{\Omega}_{\ell,\mu},
\end{align}
where $N_\text{tot}$ is the total number of sites in 3D, $\ell=e,o$ refers to even ($e$) and odd ($o$) layers, and $1 \le \mu \le 3$ is the sublattice index. We assume the inter-layer exchange coupling to be isotropic for simplicity. We obtain $g_2(M)$ by adopting the Y/UUD/V and umbrella states as $X$ and $X'$ in Eq.~\eqref{eq:g2} and use a third-order spline function to obtain a approximate continuous function. $g_1(M)$ is then obtained from Eq.~\eqref{eq:ansatz}, where $\Delta E$ is for the Y/UUD/V sequence [Fig.~\ref{fig:LSW}(c)].

\begin{figure}[t]
  \includegraphics[width=0.85\hsize]{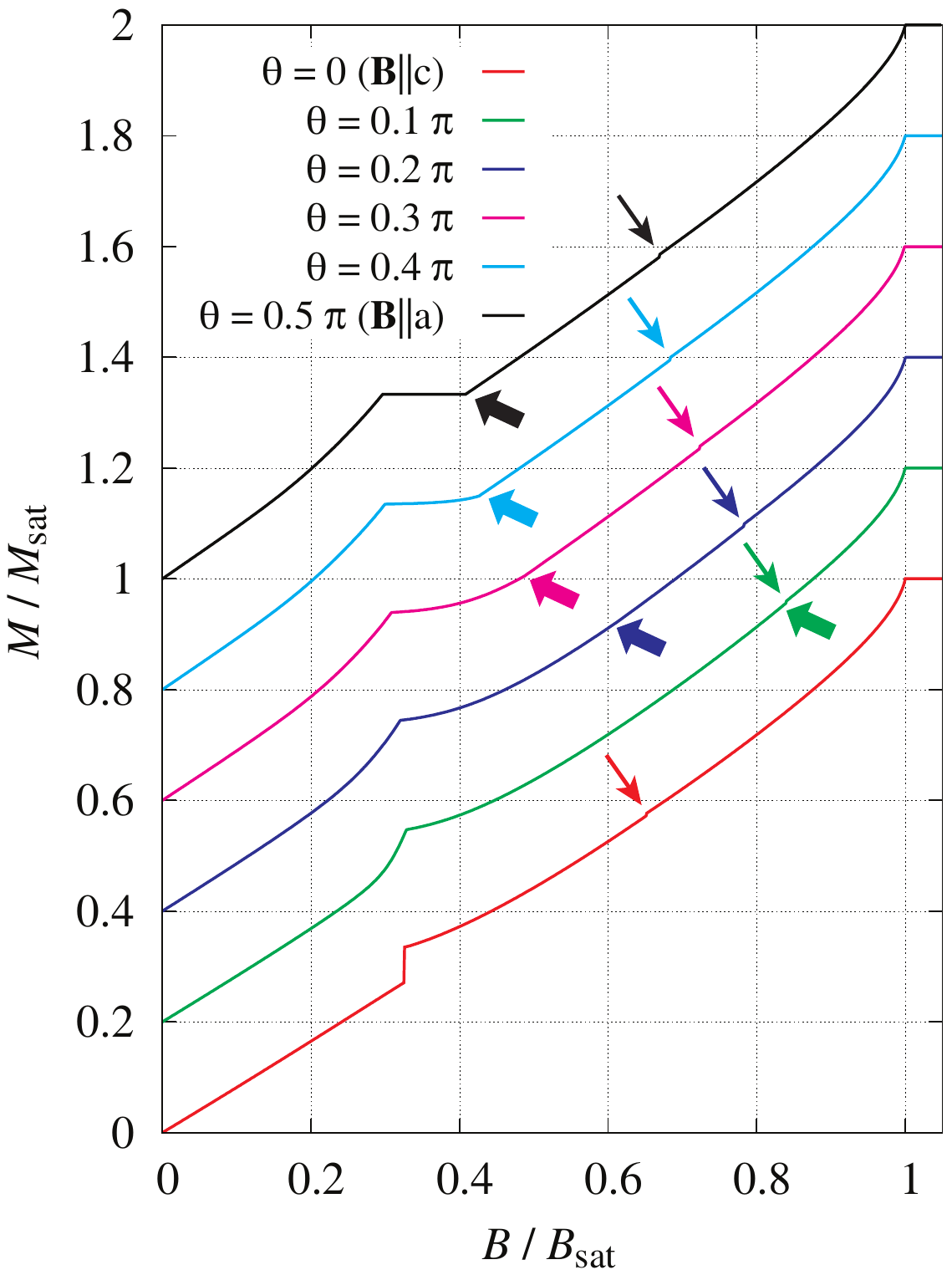}
  \caption{%
    \label{fig:MH}
    Field orientation angle ($\theta$) dependence of the magnetization curve obtained by our semi-classical mean-field theory. The horizontal axis normalization $B_\text{sat}$ depends on $\theta$.
    The arrows indicate locations of tiny jumps that can be connected to $(3/5)M_\text{sat}$ anomalies for $\mathbf{B} \parallel \hat{c}$ and $\mathbf{B} \parallel \hat{a}$.
    The thick arrows indicate a cusp associated with the LIF-UIF transition, which becomes  less evident as the field deviates from $\hat{a}$-axis and disappears for $\theta \lesssim 0.1\pi$ (see Fig.~\ref{fig:phase-diagram}).
  }
\end{figure}

\begin{figure}[t]
  \includegraphics[width=3in]{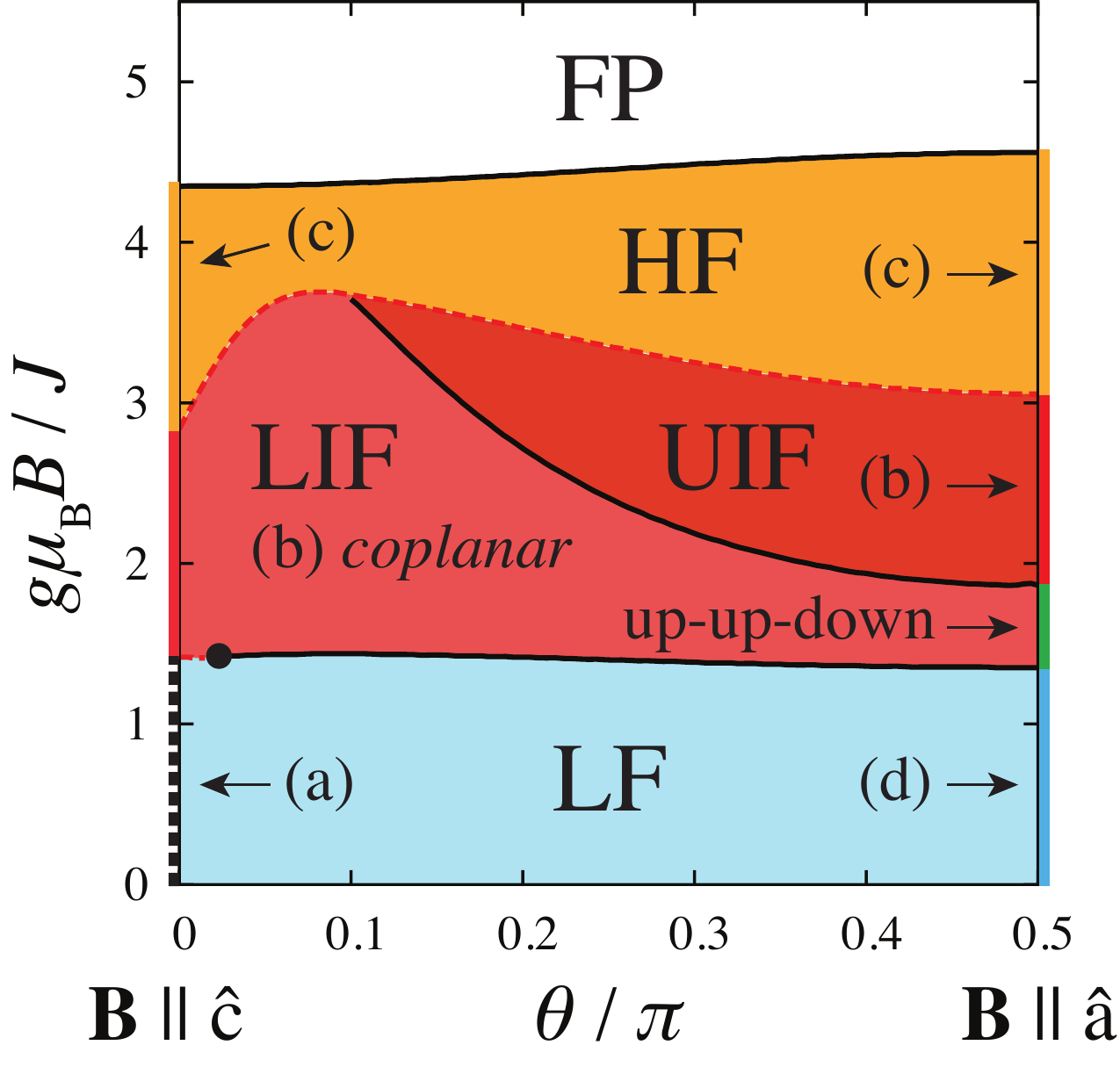}
  \caption{%
    \label{fig:phase-diagram}%
      Mean-field phase diagram parametrized by the magnetic field orientation ($\theta$) and strength ($g\mu_B B / J$). The five different phases are denoted as low-field (LF), lower intermediate-field (LIF), upper intermediate-field (UIF), high-field (HF), and fully-polarized (FP) phases. The indices (a)--(d) represent the states shown in Fig.~\ref{fig:states-3D}. The HF, UIF, and LF are non-coplanar phases with exceptions of the lines of (b)--(d) along $\theta = 0$, $\pi/2$, and the corresponding spin states are subject to deformations relative to the case with $\mathbf{B} \parallel \hat{a}$. The dashed phase boundaries are first-order transitions while the solid ones are second order transitions at the mean-field level.
  }
\end{figure}

\begin{figure}[t]
  \includegraphics[width=0.99\hsize]{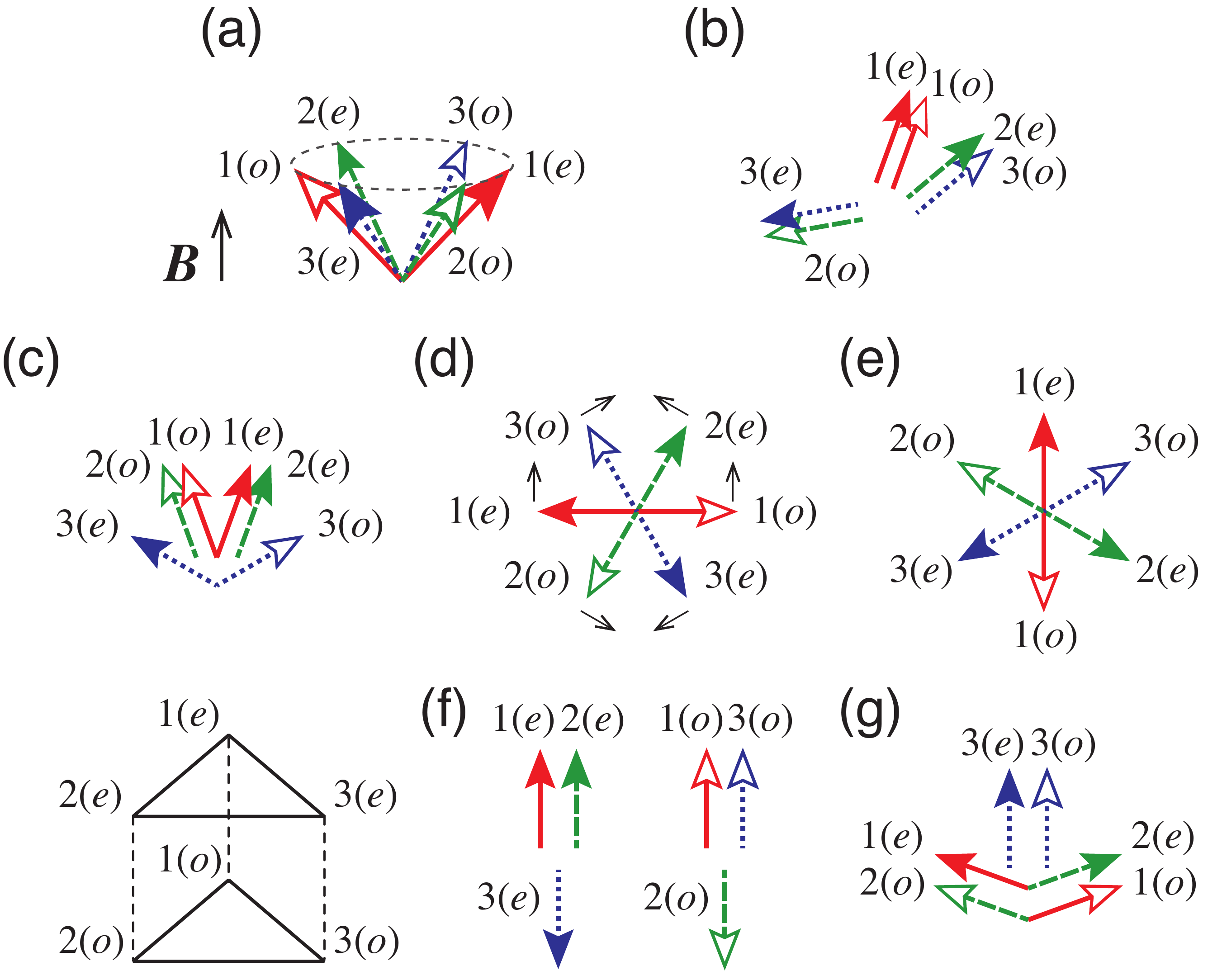}
  \caption{%
    \label{fig:states-3D}
    Magnetic states in 3D: (a) umbrella, (b) distorted V, (c) staggered V, (d) distorted combined Y, and (f) up-up-down states.
    The states (e) and (g) are not stabilized in the model described in the text. The deformation of the distorted combined Y state as a function of increasing $B$  is indicated by arrows. The non-coplanar states in the LF, UIF, and HF phases for an intermediate value of $0 < \theta < \pi/2$ are similar to (d), (b), and (c), respectively, but they are deformed because of the competition between the external magnetic field and the anisotropy.
  }
\end{figure}

\subsection{Results}
The outcome of our $T = 0$ semiclassical mean-field theory of $\ham$ is presented for $J_\parallel / J = 0.93$ and $J' / J = 0.03$. These values are almost equal to the reported ones in the literature~\cite{Susuki:2013} and they are chosen to  reproduce the magnetization curves of \baco\ for $\mathbf{B} \parallel \hat{c}$ and $\mathbf{B} \parallel \hat{a}$ (Fig.~\ref{fig:MH}).
Figure~\ref{fig:phase-diagram} shows the resulting phase diagram. At zero field, we find the 120$^\circ$ easy-plane spin configuration where the spins are anti-aligned on adjacent layers. Below, we describe the model response to a magnetic field for $\mathbf{B} \parallel \hat{c}$ and $\mathbf{B} \parallel \hat{a}$. Then, we discuss how these extreme cases are connected as a function of the angle $\theta$ between $\mathbf{B}$ and $\hat{c}$,  and we close the theory section with a discussion of the $g$-tensor anisotropy.

\subsubsection{Mean-field results for $\mathbf{B} \parallel \hat{c}$}
An infinitesimally small $\mathbf{B} \parallel \hat{c}$ induces a 3D extension of the non-coplanar umbrella state. The $xy$-components remain in a 120$^\circ$ configuration  and anti-aligned along the $c$-direction, but there is also a field-induced uniform $z$-component  [Fig.~\ref{fig:states-3D}(a)]. While this coincides with the classical ground state for $J_\parallel \le J$ in $d=3$, it is still interesting to note that a spin-wave analysis in Ref.~\onlinecite{Gekht:1997} suggests that if $J_\parallel = J$, the umbrella state is not the ground state for $J'/J \lesssim 0.1 S^{-1}$. Thus, even a rather small easy-plane anisotropy has a strong effect in the low-field regime of TLHAF.

Our analysis predicts a first order transition to a coplanar ``distorted V'' state [Fig.~\ref{fig:states-3D}(b)] at $B \simeq B_\text{sat}/3$, accompanied by a small jump in $M(B)$ (see Fig.~\ref{fig:MH}). We speculate that the cusp, instead of the small jump, reported in \baco\ at $T=1.3$K~\cite{Susuki:2013} is induced by small perturbations (such as impurities, a finite-$T$ effect, or a combination thereof). This coplanar phase is similar to the V phase that emerges {\it above} the UUD phase  in the 2D isotropic case; the rather small easy-plane anisotropy destabilizes the collinear UUD phase predicted for the isotropic model with $J'/J \lesssim 0.1 S^{-1}$.\cite{Gekht:1997}

\begin{figure}[t]
  \includegraphics[width=2.8in]{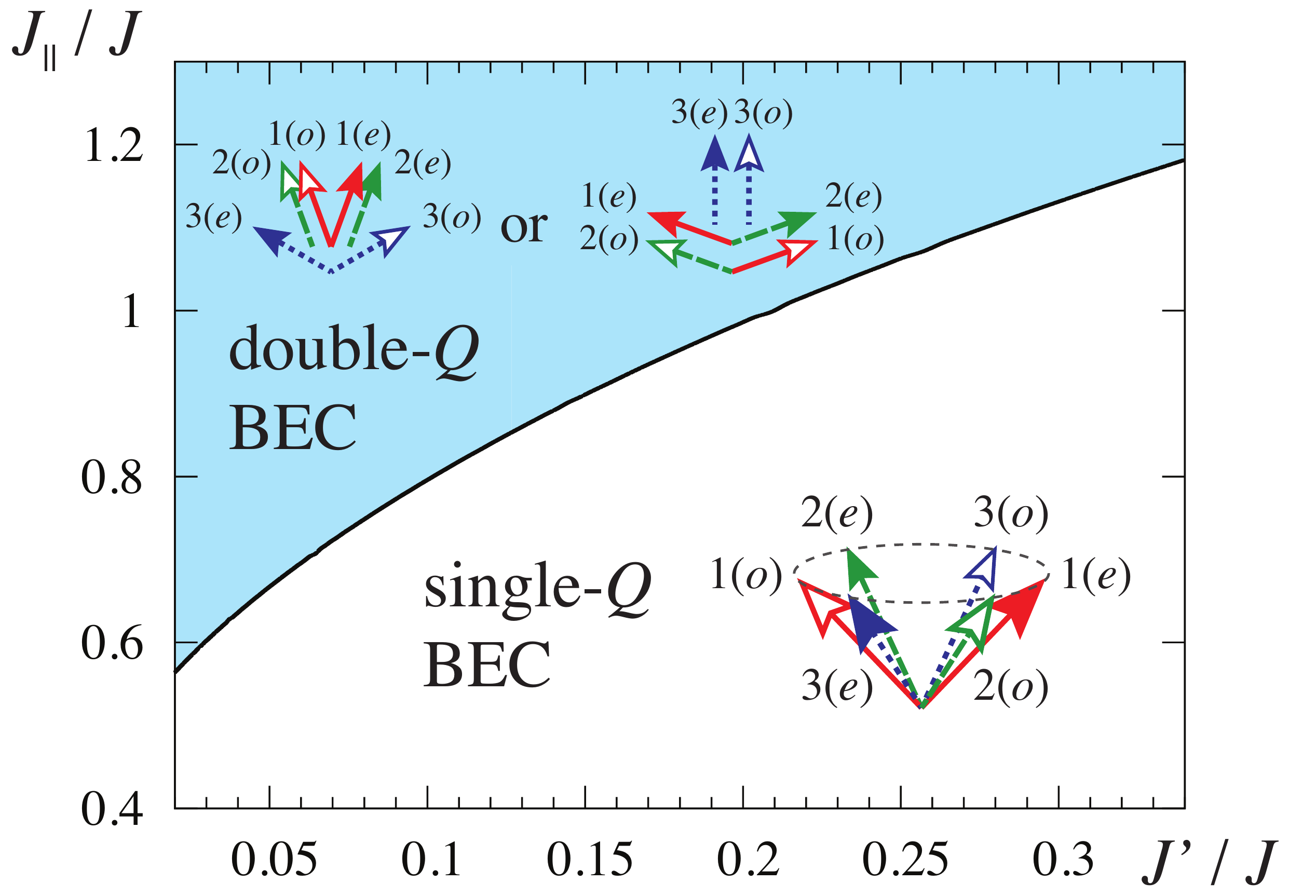}
  \caption{%
    \label{fig:BEC}
    Ground-state phase diagram for $B = B_\text{sat}$ and $\mathbf{B} \parallel \hat{c}$ obtained by the dilute-gas analysis.
  }
\end{figure}

The in-plane spin components of even and odd layers are not anti-aligned for every sublattice ($1 \le \mu \le 3$) in the distorted V state.
For an XY-like spin system, like the one under consideration, it is very plausible that a state with a staggered  transverse spin component  along the $\hat{c}$-axis (${\bf Q} \cdot \hat{c} = \pi$) is realized close enough to the saturation field. This is a consequence of the Bose-Einstein condensation (BEC) of the lowest-energy magnon ($Q_z = \pi$) of the FP state expected at $B = B_\text{sat}$.\cite{Giamarchi1999coupled,Giamarchi2008bose,Nikuni1995Hexagonal} In conformity with this expectation, we observe a first-order transition to a different coplanar ``staggered V'' state [Fig.~\ref{fig:states-3D}(c)] that persists up to $B=B_\text{sat}$. When the critical point is approached from $B > B_\text{sat}$, this state corresponds to a double-$\mathbf{Q}$ magnon condensate, i.e., a BEC in a single particle state that is a linear combination of the $\mathbf{Q} = (4\pi/3,0,\pi)$ and $-\mathbf{Q}$ states with a particular relative phase.\cite{Nikuni1995Hexagonal} This result is consistent with a dilute-gas analysis of the magnon BEC,\cite{Giamarchi1999coupled,Giamarchi2008bose,Nikuni1995Hexagonal} which predicts the double-{\bf Q} BEC leading to either the staggered V state or a different state shown in Fig.~\ref{fig:states-3D}(g) for relatively small $J'/J$ and large $J_\parallel / J$ (otherwise the single-{\bf Q} BEC leading to the umbrella state is predicted), as is shown in Fig.~\ref{fig:BEC}.
The transition between the distorted V and the staggered V states implies a discontinuous change of the relative orientation between the even and odd layer spins. As is clear from the above argument, the inter-layer coupling is responsible for this outcome. This transition is accompanied by an almost negligible jump in $M(B)$ at $M \approx (3/5)M_\text{sat}$ (Fig.~\ref{fig:MH}). A similar anomaly for $\mathbf{B} \parallel \hat{a}$ (see below) is continuously connected with the one under consideration  as the field orientation is rotated from $\hat{c}$ to $\hat{a}$.
Experimentally, such an anomaly at $M \approx (3/5)M_\text{sat}$ has been detected in \baco\ only for $\mathbf{B} \parallel \hat{a}$. \cite{Susuki:2013}  We speculate that the predicted anomaly for $\mathbf{B} \parallel \hat{c}$ could be beyond the experimental resolution of bulk magnetization measurements, but it should be possible to test this prediction by NMR measurements (see, however, the discussion in \ref{subsec:H||c,large-H}).

Thus, for $\mathbf{B}\parallel\hat{c}$, the easy-plane anisotropy dominates the low-field regime, $B \lesssim B_\text{sat}/3$, stabilizing the umbrella state and destabilizing the UUD phase.  The sequence of phases for $B \gtrsim B_\text{sat}/3$ is nearly identical to that of the quasi-2D quantum isotropic model.\cite{Gekht:1997} 

\begin{figure*}
  \includegraphics[width=0.95\hsize]{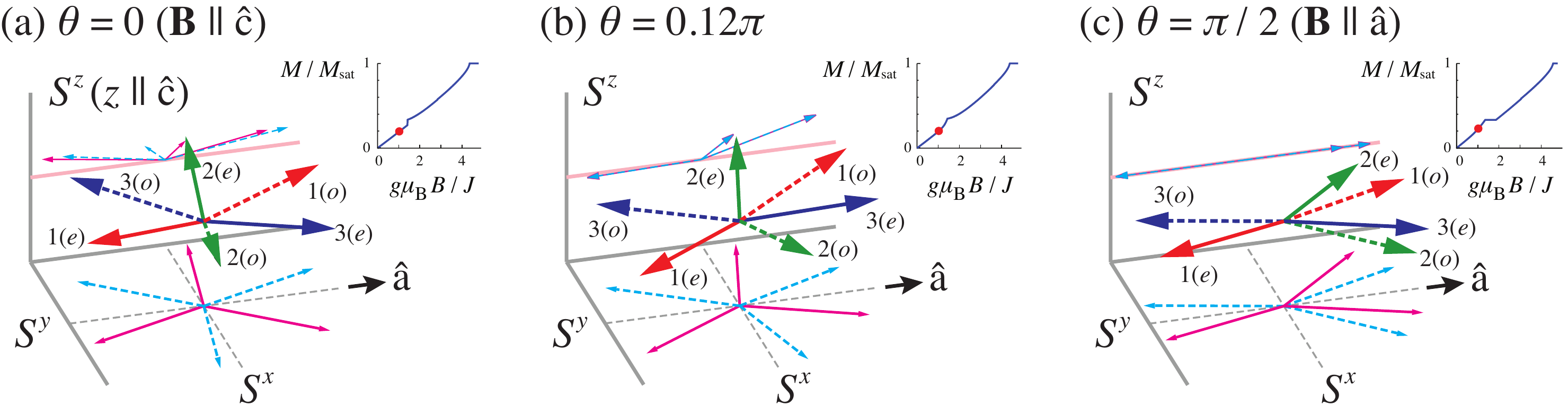}
  \caption{%
    \label{fig:LFC}%
    Sublattice spin configuration in the low-field phase ($g \mu_B B / J = 1.0267$) for different external field orientations, as calculated by the mean-field model described in the text: (a) the umbrella state, (c) the distorted combined Y state, and (b) an intervening state between these states.  Projections on the $xy$ and $xz$ planes are also shown. The insets show the corresponding points in the $M(B)$ curve.
}
\end{figure*}

\subsubsection{Mean-field results for $\mathbf{B} \parallel \hat{a}$}
Next, we consider the case $\mathbf{B} \parallel \hat{a}$. Because the field is parallel to the easy-plane, the spins are always parallel to the $ab$-plane.  This seems to be the only relevant effect of the easy-plane anisotropy at the mean-field level.
For arbitrary small field, the ground state is the ``distorted combined Y'' state shown in Fig.~\ref{fig:states-3D}(d). This state is continuously connected  to a particular orientation of the 120$^\circ$ configuration for $B\to0$.
At the classical level, this state is quasi-degenerate with the state shown in Fig.~\ref{fig:states-3D}(e). The classical energy difference is $\sim (\mu_B B/J)^6$. Quantum fluctuations further stabilize the combined distorted Y state. An  increasing magnetic field deforms of the combined distorted Y state in the $ab$-plane. This deformation is smoothly connected to the UUD state [Fig.~\ref{fig:states-3D}(f)], resulting in a $M_\text{sat}/3$ plateau.

On further increasing the magnetic field, the UUD phase undergoes a second-order phase transition to a high field distorted V phase [Fig.~\ref{fig:states-3D}(b)]. The spin configuration is identical to what we already discussed above for $\mathbf{B} \parallel \hat{c}$, but in this case the spins are confined to the $ab$-plane. Like in the case of the UUD state,\cite{Chubukov91} quantum fluctuations play an essential role in stabilizing this state:  the classical limit of the weakly-coupled triangular layers of the XY antiferromagnet leads to a different magnetic ordering shown in Fig.~\ref{fig:states-3D}(g) in the high-field regime.\cite{Plumer1990Magnetic}
The exchange anisotropy is also believed to play an important role in stabilizing these quantum states because $J_\parallel = J$ leads to a gapless Goldstone mode in these phases. Finally, as for $\mathbf{B} \parallel \hat{c}$, the model predicts a first-order phase transition between the distorted V and staggered V states. As mentioned above, this transition is accompanied by a tiny jump in $M(B)$ at around $M = (3/5)M_\text{sat}$ (see Fig.~\ref{fig:MH}), which may correspond to the anomaly observed in magnetization measurements.~\cite{Susuki:2013}

\subsubsection{Intermediate field orientation $0 < \theta < \pi/2$}
Figure~\ref{fig:LFC} shows the spin configuration in the successive phases as the field orientation varies in the low-field regime $B \lesssim B_\text{sat}/3$.  A rotation of the applied field ($\theta = 0 \to \pi/2$) causes a continuous distortion between the (non-coplanar) umbrella and the (coplanar) distorted combined Y states, which corresponds to the  low-field (LF) phase shown in Fig.~\ref{fig:phase-diagram}. The non-coplanar state highlighted in the middle panel of Fig.~\ref{fig:LFC} is similar to the one corresponding to the experimental results presented in Fig.~\ref{fig:BcSpectra}, i.e. $\theta=15^\circ$ (see below).

Similarly, rotating the field in the UUD phase at $B \approx B_\text{sat}/3$ for $\theta = \pi/2$ corresponds to adding a transverse field component to the UUD state. Consequently, the UUD state is continuously deformed  into a distorted V state, which is smoothly connected to the state that we discussed for $\theta = 0$ ($\mathbf{B} \parallel \hat{c}$). We refer to this state as the ``lower intermediate-field'' (LIF) phase.
For $0 < \theta < \pi/2$, the spins are on the plane subtended by $\mathbf{B}$ and $\hat{c}$. For $\theta = 0$ ($\mathbf{B} \parallel \hat{c}$), the U(1) symmetry of  spin rotations along the $\hat{c}$-axis is spontaneously broken implying that the spins are on the plane subtended by $\mathbf{B}$ and an arbitrary axis perpendicular to the  $\hat{c}$-direction.

The distorted V state stabilized for $B \gtrsim B_\text{sat}/3$ for $\theta = \pi/2$ ($\mathbf{B}\parallel\hat{a}$) does not belong to the LIF phase, but to a different phase that we call the ``upper intermediate-field'' (UIF) phase. To see this subtle but crucial difference, we recall that in this case the distorted V state lies on the $ab$-plane, which is perpendicular to the spin plane in the LIF phase. In addition, the rotation of the field orientation from $\theta = \pi/2$ causes a \textit{non-coplanar} deformation of its spin configuration in contrast to the UIF phase.
Thus, at the mean field level, the LIF-UIF second order transition signaled by a cusp  in the magnetization curve (Fig.~\ref{fig:MH})  is associated with the appearance of finite scalar spin chirality.
The UIF phase does not extend up to $\theta = 0$ but terminates in the middle (Fig.~\ref{fig:phase-diagram}).
In the high field regime, the staggered V states for $\mathbf{B} \parallel {\hat{a}}$ and $\mathbf{B} \parallel {\hat{c}}$ are continuously connected through the field rotation. This phase is dubbed ``high-field'' (HF) phase.

\subsection{ $g$-factor anisotropy }
So far, we have analyzed the easy-plane XXZ Hamiltonian based on a mean-field approach as an effective model for the spin-1/2 moments in \baco, assuming the isotropic $g$-tensor. Below, we discuss  its possible anisotropy.

The easy-plane exchange anisotropy implies that it is easier to polarize spins along the ``hard'' axis (i.e., the $\hat{c}$-axis).
The lower energy cost of a uniform magnetization component along the $\hat{c}$-axis  (the $S^z$-$S^z$ antiferromagnetic coupling  is smaller than  the $S^x$-$S^x$ coupling) leads to a lower saturation field along this axis.
Indeed, a simple microscopic analysis shows that the Zeeman energy required to fully polarize the spins is
$h_\text{sat,3D}^{\perp} \equiv g_{\perp} \mu_B B_\text{sat,3D}^{\perp} = (9J + 4J')S$ for $\mathbf{B}\perp\hat{c}$
and
$h_\text{sat,3D}^{\parallel} \equiv g_{\parallel} \mu_B B_\text{sat,3D}^{\parallel} = (3J + 6J_\parallel + 4J')S$ for $\mathbf{B}\parallel\hat{c}$.
Consequently,
\begin{align}
  h_\text{sat,3D}^{\perp} > h_\text{sat,3D}^{\parallel}~~\text{if  $J > J_\parallel$.}
\end{align}
This feature is captured in the phase diagram (Fig.~\ref{fig:phase-diagram}) obtained by our mean-field theory, which reproduces the exact saturation field for $\mathbf{B}\perp\hat{c}$ and gives an excellent approximation for $\mathbf{B} \parallel {\hat{c}}$.
By assuming $J_\parallel / J = 0.93$ and $J'/J = 0.03$, we obtain
\begin{align}
  g_{\parallel} / g_{\perp} \approx 0.93
\end{align}
as a rough estimate based on the measured values of the saturation fields $B_\text{sat,3D}^{\parallel} = 32.8$T and $B_\text{sat,3D}^{\perp} = 31.9$T.\cite{Susuki:2013}
We also point out that $g_\parallel < g_\perp$ is a rather generic property expected for the pseudospin-1/2 of the Kramers doublets in Co$^{2+}$ ions with easy-plane exchange anisotropy.~\cite{Palii2003} The  second-order perturbation theory presented in Ref.~\onlinecite{Palii2003} leads to $g_{\parallel} / g_{\perp} \approx 0.97$ by assuming a realistic set of parameters for the Co$^{2+}$ ion.

In contrast, EPR measurements support the opposite inequality: $g_{\parallel} = 3.87$ and $g_{\perp} = 3.84$ ($g_{\parallel} / g_{\perp} \approx 1.01$).\cite{Susuki:2013}
One may argue, however, that there could be a dynamical shift of the measured $g$-values because of short-range ordering effects that should still be present at  the temperature of the EPR measurements, $T=20$K , which is comparable to $J$. EPR measurements at high enough $T$ or $B$ will be of great help to settle this issue.

\section{Experimental Details}
The measurements were carried out on a 30mg single crystal of \baco\ synthesized by the traveling-solvent floating-zone method, as for previous thermal and neutron scattering experiments \cite{Zhou:2012}. Approximate dimensions were 4mm$^2$ $\times$ 1.2mm thickness. Experiments took place in two laboratories, depending on the magnetic field strength. Fields up to 12T were available in the UCLA laboratory, and experiments at higher fields (14.5-30T) were performed in a resistive Bitter magnet at the National High Magnetic Field Laboratory in Tallahassee, FL.

For the lower field measurements, performed at UCLA, the sample and NMR coil were mounted on a single axis piezo-driven rotator (Attocube ANRv51/RES). Relative rotation angle was determined using a resistive sensor built into the rotator, and complementary absolute information was obtained by observing the angle variation of the NMR transition frequencies.\cite{Appendix2} For most of these measurements, the rotation axis was lying 15$^\circ$ out of the $ab$-plane; consequently, the angle closest to ${\mathbf B}\parallel\hat{c}$ was also $\theta_{min}=15^\circ$. The high-field probe was equipped with a mechanically-driven goniometer. For those measurements, made at the NHMFL, the rotation axis was in-plane to within $\pm3^\circ$.

There are two crystalographically inequivalent Ba sites, hereafter referred to as Ba(1), Ba(2), both with uniaxial symmetry (see Fig. \ref{fig:xtal} in the Appendix). The Ba(1) location is between Co ions of adjacent layers, at (0,0,1/4) and (0,0,3/4). The Ba(2) site is equidistant from three Co ions in a layer, but offset 1.3\AA$\parallel \hat{c}$.\cite{Appendix2} Since the isotopic concentrations are 11\% and 6\% for $^{137}$Ba and $^{135}$Ba, respectively, we mostly confined our measurements to the more abundant species, for which the gyromagnetic ratio is $^{137}\gamma/2\pi=4.73158$MHz/T. The nuclear spin is $^{137}I=3/2$, so there are two satellite transitions $\left<\pm 3/2 \leftrightarrow\pm 1/2\right>$ in addition to the central $\left<+1/2 \leftrightarrow -1/2\right>$. The detailed measurements targeted the latter. The inequivalent magnetic environments developing in the ordered phases were used to infer qualitative and quantitative information about the ordered moments.

The local field at the nuclear sites can be written as $B_{local}=B_{macro}+B_{hf}+B_o$, with $B_{macro}$ the macroscopic field arising from applied, Lorentz, and demagnetization sources, $B_{hf}$ the hyperfine field, and $B_o$ the orbital contribution. Labeling of the various NMR transitions and extraction of the NMR parameters are outlined in the Appendix. The spectra that follow are constructed from frequency swept Fourier transform sums. They are all presented in terms of frequency shifts originating with the Ba(1) hyperfine fields, $\Delta\nu=$ $^{137}\gamma B_{hf}$; to within experimental uncertainties, contributions from the orbital and macroscopic fields are removed, as is the frequency shift originating with electric quadrupole coupling to the lattice electric field gradient.

The crystal orientations were established by inspection of the resonance frequency of the different quadrupolar transitions for the Ba sites and isotopes upon rotation of the applied field about the \ita-axis, with an example presented in the Appendix. As discussed in detail below, our NMR results unambiguously indicate that the UUD phase occurs for $\mathbf{B}\perp \hat{c}$, consistent with the findings of Ref.~\onlinecite{Susuki:2013} but opposite to those of Ref.~\onlinecite{Zhou:2012}. Moreover, all our measurements replicate precisely the phase transition lines described in  Ref.~\onlinecite{Zhou:2012}, but for $\hat{a}\leftrightarrow\hat{c}$. For this reason, in what follows, the phase diagram shown in  Ref.~\onlinecite{Zhou:2012} has been adopted assuming $\hat{a}\leftrightarrow\hat{c}$.

\section{Experimental Results and Discussion}
\subsection{NMR for ${\mathbf B}\perp\hat{c}$}
\subsubsection{Low-field Measurements ($B<12\rm{T}$) }

The NMR spectrum of $I=3/2$ Ba nuclei in equivalent environments comprises three spectral lines, which correspond to the nuclear transitions $\left<m_I\leftrightarrow m_I-1\right>$ with $m_I=\pm 3/2, \pm 1/2$. When the magnetic interaction dominates over the electric quadrupole interaction, the resonance frequency of the central nuclear transition is approximated up to second order in the quadrupole interaction as \cite{Appendix2}
\begin{align}\label{eq:NMR_f}
\nu_{\frac{1}{2}\leftrightarrow-\frac{1}{2}}
=&\gamma \left(B_{macro}+B_{orb}+B_{hf}\right)
\notag\\
  &+\frac{3\nu_Q^2}{16\gamma B}\left(1-\cos^2\theta \right)\left(1-9\cos^2\theta \right) ,
\end{align}
where $B_{macro}$ is the macroscopic field in the crystal interior, consisting of applied, Lorentz and demagnetization fields. Below, this sum is written simply as $B$. $B_{orb}$ is the local Ba(1) orbital field, and similarly $B_{hf}$ is the hyperfine field. $\nu_Q$ is the nuclear quadrupole frequency, also for the Ba(1) site. In the paramagnetic state and where there is a well-defined magnetic spin susceptibility $\chi_s$, the hyperfine field can be written as $B_{hf} \equiv K_s B_{0}$, and the hyperfine shift $K_s=A\chi_s$.\cite{Appendix2} As mentioned above, the presented spectra are referenced to the hyperfine fields at the Ba(1) site, with $\Delta\nu\equiv$ $^{137}\gamma B_{hf}$, with the macroscopic, orbital, and quadrupolar contributions subtracted out. Note that the sample geometry leads to variations in the demagnetization field through the crystal, in turn contributing to line broadening of order of tens of kilohertz in the typical fields discussed here.

\begin{figure}[h]
\includegraphics[width=3.3in]{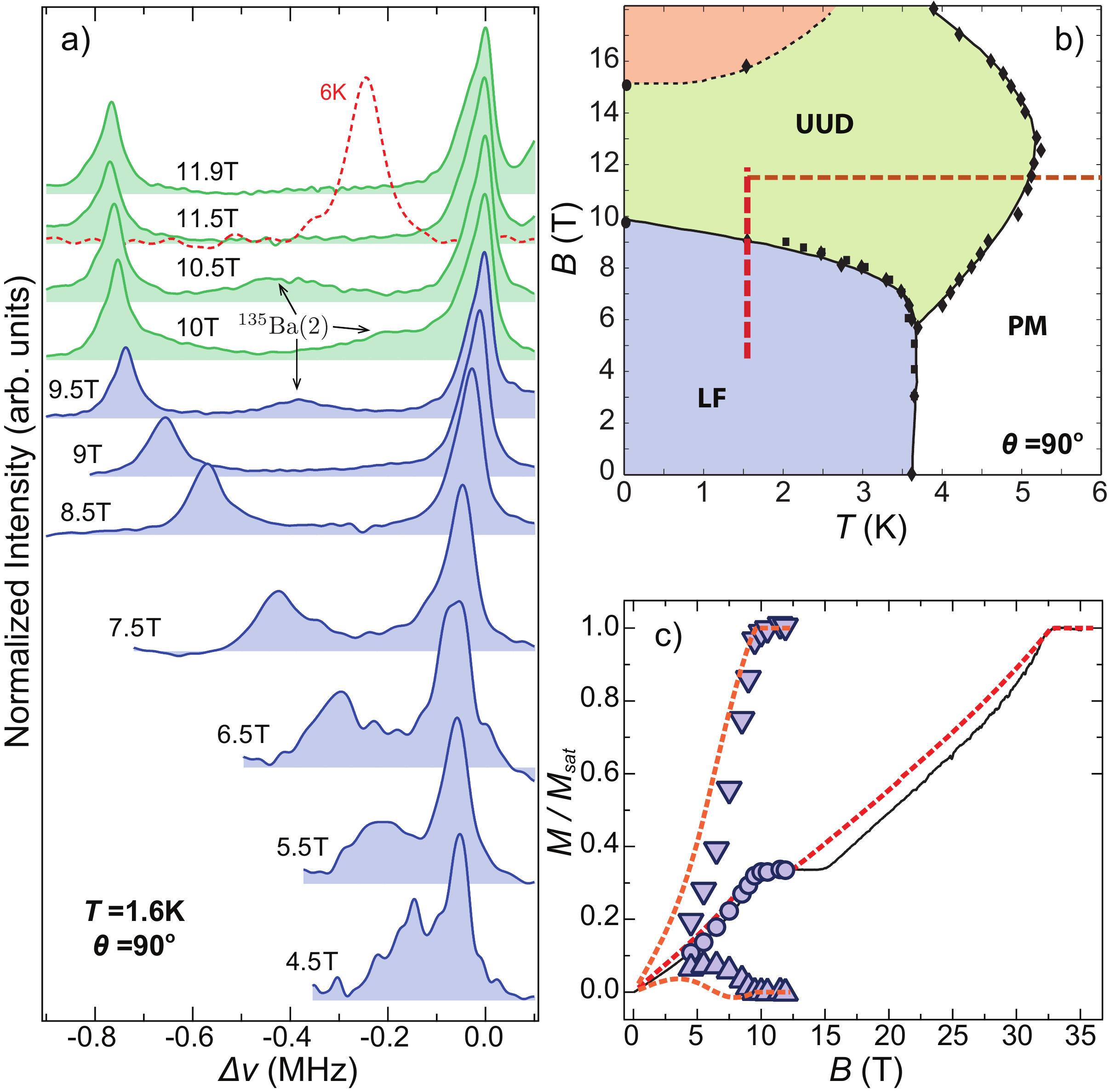}
\caption{(a) Field evolution of the $^{137}$Ba(1) hyperfine frequency shifts (see text) for $\theta=90^{\circ}$ at $T=1.6$K. The dashed line shows the paramagnetic phase spectrum at $T=6$K for comparison. (b) Experimental phase diagram for ${\mathbf B}\perp\hat{c}$, after Ref.~\onlinecite{Zhou:2012}. The vertical dashed line corresponds to the field range and temperature in (a). (c) Magnetization vs.~magnetic field. The solid, dashed lines are from magnetization results, and the model of Eq. \ref{eq:MF}, respectively. The data points are derived from the NMR spectra: circles are the first moment of the full spectrum, properly normalized, and the triangles are associated to the hyperfine shifts of the two Ba(1) local environments. }
\label{fig:BabSpectra}
\end{figure}

In Fig.~\ref{fig:BabSpectra}a, $^{137}$Ba(1) spectra for the central nuclear transition are shown, recorded for $\theta=90^{\circ}$ at $T=1.6$K and fields 4.5--12T. These data correspond to the cut through the $B$-$T$ phase diagram of Fig.~\ref{fig:BabSpectra}b illustrated by the heavy (vertical) dashed line. For completeness, we have included the boundary lines inferred from magnetization and specific heat studies described in Ref.~\onlinecite{Zhou:2012}. On increasing the magnetic field $B$, a spectral line-splitting (doubling) develops; the features define two distinct local field environments. For $B\gtrsim10$T, the resonance frequency difference \mbox{$\Delta f=$ $^{137}\gamma \Delta B_{hf}$} is constant, indicating no further changes to the hyperfine fields.

Identification of the fixed line splitting for $B\gtrsim10$T as the UUD phase is straightforward. In the magnetically ordered states, the hyperfine field at each Ba(1) nuclear site stems predominantly from coupling to the moments associated with its two nearest-neighbor (NN) Co ions, located at relative positions $\mathbf{r}=(0,0,\pm c/4)$, i.e. on adjacent even/odd layers \cite{Appendix1}. Hence, at the Ba(1) sites we have $\mathbf{B}_{hf}=\sum_{i} \mathbb{A}_i\cdot\boldsymbol{\mu}_i$, where $i=e,o$, $\boldsymbol{\mu}_i$ is the moment of i-th Co site, and $\mathbb{A}_i$ is the hyperfine coupling tensor between the Ba(1) and the i-th Co site. The analysis is simplified after taking into account this local symmetry of the Ba(1) site, which leads to only diagonal non-zero elements. In the case of the UUD phase, where $\boldsymbol{\mu}_i=(\pm M_\text{sat},0,0)$, two different Ba(1) environments are possible: ($\uparrow\uparrow$), ($\uparrow\downarrow$), with 1:2 intensity, respectively (see Fig.~\ref{fig:states-3D}(f)). The first configuration leads to a hyperfine frequency shift $\Delta\nu\equiv\gamma B_{hf}^{\uparrow\uparrow}$ corresponding to the full Co moment on the NN sites, and the second has vanishing shift. The local magnetization for the distinct local environments is calculated from the two NMR peaks in Fig.~\ref{fig:BabSpectra}a as $M_{loc}=\Delta\nu/\gamma A_{\perp c}$, where $A_{\perp c}$ is the hyperfine coupling constant for the particular field orientation.\cite{Appendix2} The result is shown in Fig.~\ref{fig:BabSpectra}c, which also includes the calculated local magnetization corresponding to the spectral first moment that scales quantitatively with the average moment/Co. Evidently, the first moment of the full absorption spectrum for the Ba(1) central transition follows the average magnetization previously measured (solid line), and this is reproduced well by the model results (dashed lines).

\begin{figure}[h]
\includegraphics[width=2.65in]{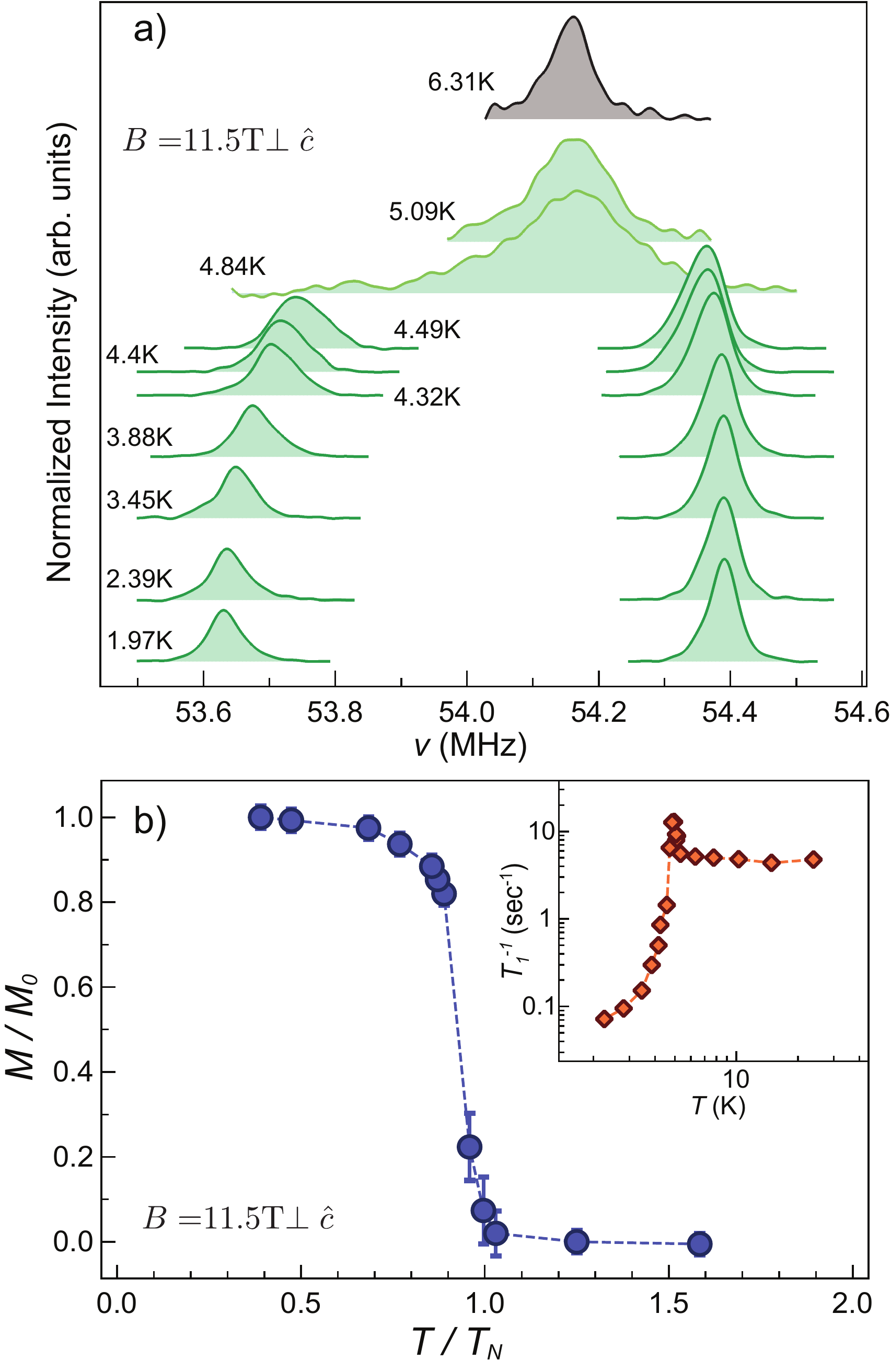}
\caption{(a) NMR spectra of the $^{137}$Ba(1) $\left<1/2\leftrightarrow -1/2\right>$ transition for different temperature values at $\mathbf{B}$=11.5T$\perp\hat{c}$. (b) Temperature dependence of the ordered moment amplitude in the UUD phase. \textit{Inset:} Spin-lattice relaxation rate \textit{vs.} temperature for $\mathbf{B}$=11.5T$\perp\hat{c}$.}
\label{fig:UUD_vsT}
\end{figure}
The temperature dependence of the ordered moment amplitude in the UUD phase was measured at $B=11.5$T, as denoted by the horizontal cut (dashed line) through the phase diagram in Fig.~\ref{fig:BabSpectra}b. The results are summarized in Fig.~\ref{fig:UUD_vsT}, which shows the relevant spectra, followed by the scaled hyperfine frequency splitting between the two Ba(1) local environments {\it vs.}~temperature. As expected for a plateau phase, there is little variation at lower temperatures, but it drops suddenly for $T>5$K. The relaxation rate is roughly independent of temperature in the paramagnetic state, which is the expected result where antiferromagnetic correlations are not particularly significant. Below the ordering temperature, 1$/T_1$ drops abruptly following a critical enhancement close to $T_N$. Further measurements will be necessary to establish whether this is a first-order, or continuous transition. The expected result would correspond to $Z_3 \times Z_2$ symmetry breaking for a system with an antiferromagnetic interlayer coupling, for which the thermal transition is of the second order.\cite{Oshikawa2000}
For the moment, we comment only that the onset coincides with the thermodynamic results.\cite{Zhou:2012}

It should be noted that a slightly reduced moment in the UUD phase relative to the saturated value $M_\text{sat}$ was observed for the large-spin ($S=5/2$) TLAF RbFe(MoO$_4$)$_2$.\cite{Svistov:2006} Based exclusively on the uncertainties of the results reported here, a similar reduction cannot be ruled out. 

Finally, the monotonically reduced hyperfine splitting for $B<B_{c1\perp}\simeq10$T indicates a continuous transition from the low-field coplanar (LF) phase. The NMR spectra in the LF phase (blue in Fig.~\ref{fig:BabSpectra}a) are consistent with in-plane spin configurations related by a reflection symmetry about the field direction in adjacent layers, in agreement with the distorted combined Y state [Fig.~\ref{fig:states-3D}(d)] predicted by our model. 
Other, slightly different proposals for this phase~\cite{Zhou:2012,Susuki:2013} (specifically, the Y state favored in the 2D limit) can be ruled out, at least for $T \gtrsim 1.5\,$K.

\subsubsection{High-field Measurements ($B>14\rm{T}$) }
\begin{figure}[h]
\includegraphics[width=3.3in]{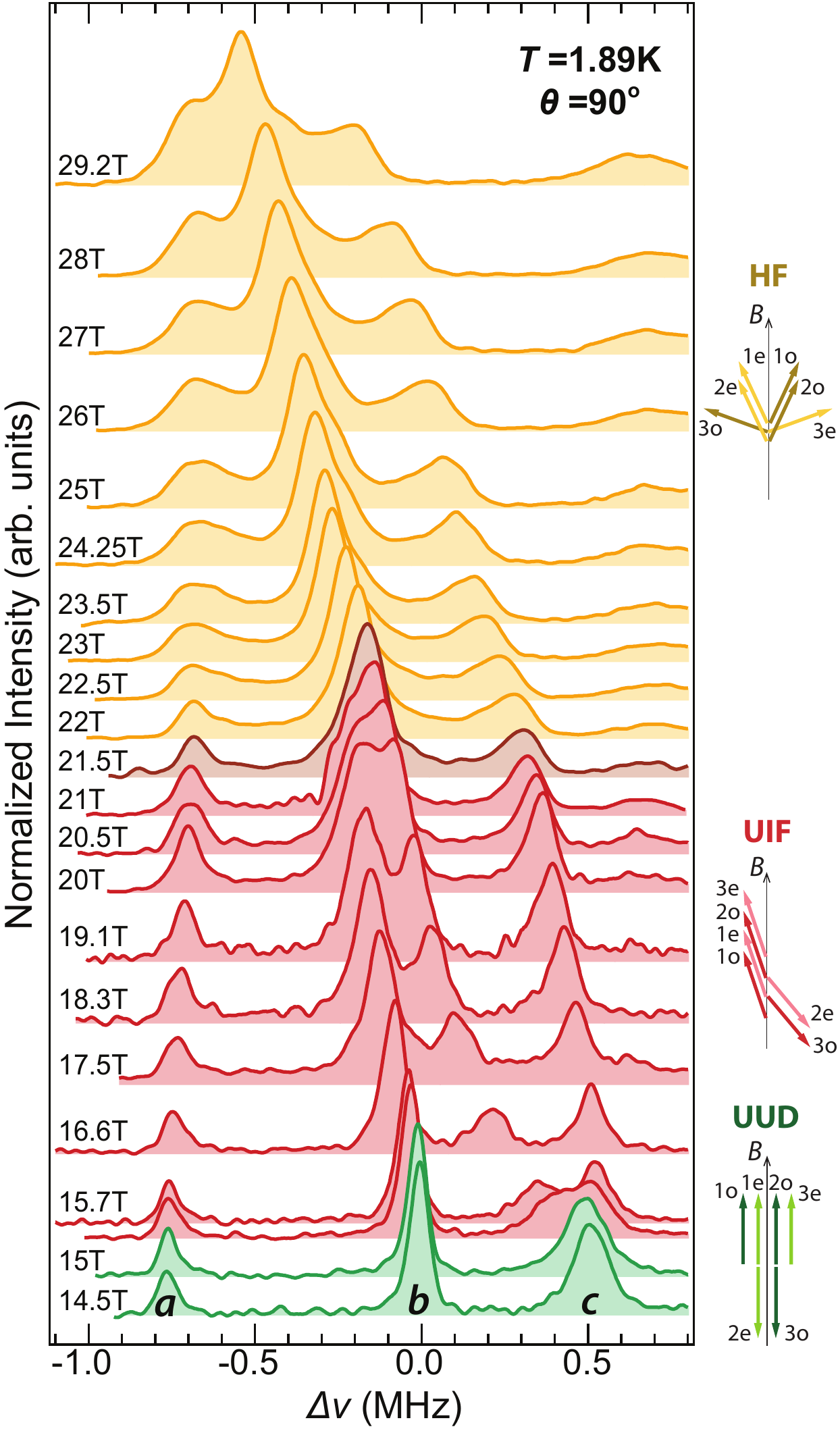}
\caption{$^{137}$Ba central transition absorption spectra at varying applied magnetic fields for $\mathbf{B}\perp\hat{c}$. Peaks \textit{a,b,c} comprise Ba(1) and Ba(2) signal as discussed in the text. Spectra recorded in each of the accessed phases is distinguished by color, with the baseline offset according to the field strength. The phase transitions inferred to be UUD$\to$UIF and UIF$\to$HF are identified from the NMR spectrum evolution, as well as spin lattice relaxation measurements (Fig.\ref{fig:T1_vsBpara}). The ordered moment configuration in each phase is sketched on the right side, with the three sublattices (1-3) on two adjacent triangles ($odd/even$) shown.}
\label{fig:Sp_vsB_a_HFL12}
\end{figure}

\begin{figure}[h]
\includegraphics[width=3.1in]{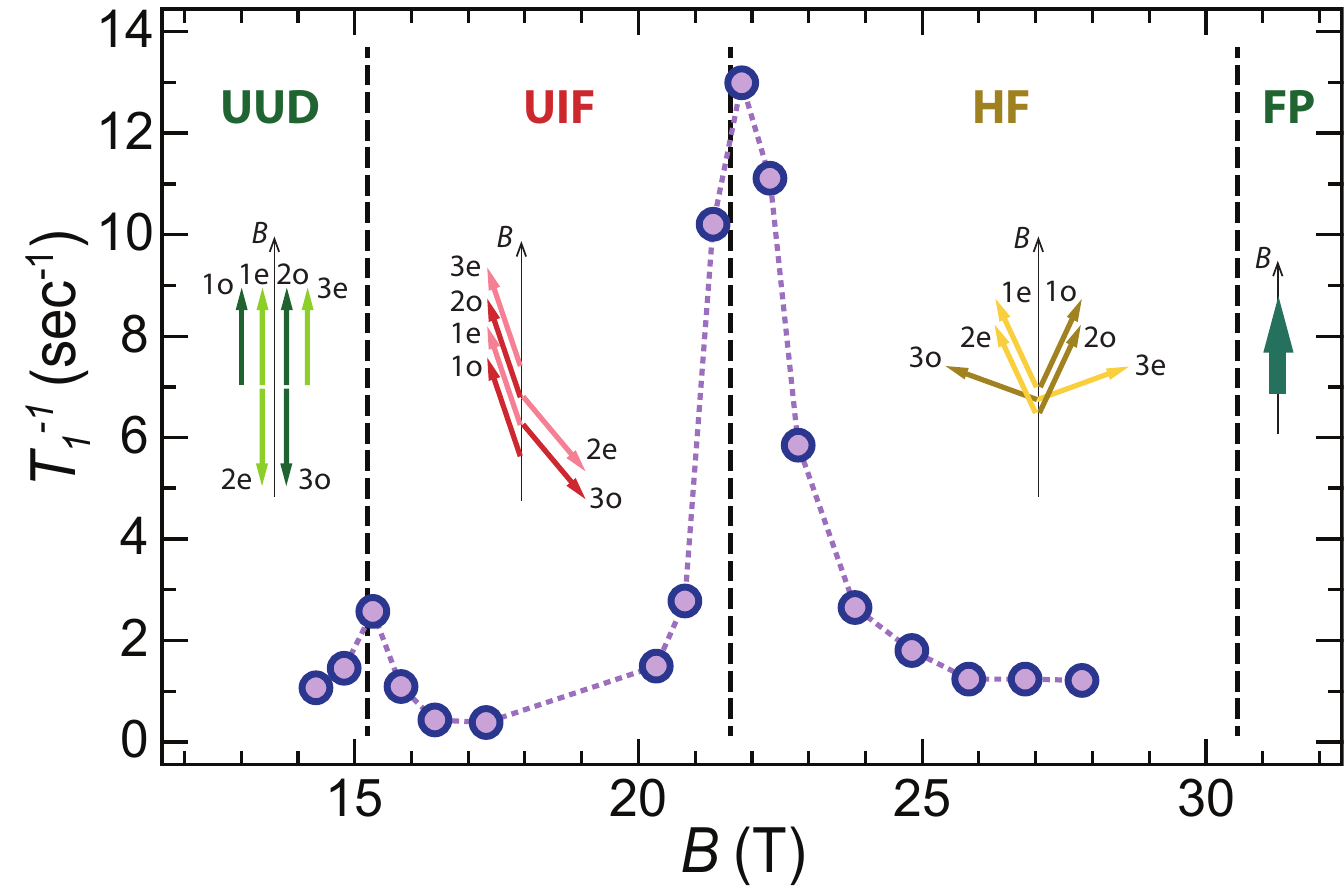}
\caption{Spin-lattice relaxation rate vs Field for $\mathbf{B}\perp\hat{c}$. The peaks near 15T and 22T are indicative of phase transitions. The low field peak is assigned to UUD$\to$UIF, and the high-field one tentatively to UIF$\to$HF. In the interest of simplicity, the measurements were isolated to the peak \textit{a} in Fig. \ref{fig:Sp_vsB_a_HFL12}, which originates with Ba(1) sites only. }
\label{fig:T1_vsBpara}
\end{figure}

The $^{137}$Ba spectra associated with the approach to the saturation field, recorded at $T\sim1.9$K, are shown in Fig. \ref{fig:Sp_vsB_a_HFL12}. The data were obtained using a dc resistive magnet at the NHMFL with maximum field 30T. The spectra cover the central transition for both Ba(1) and Ba(2) sites, and there is some overlap in each spectrum shown. The zero frequency corresponds to vanishing hyperfine field for Ba(1) sites. As the field is varied, two phase transitions are inferred from the spectral evolution (Fig. \ref{fig:Sp_vsB_a_HFL12}), as well as from relaxation measurements (Fig. \ref{fig:T1_vsBpara}): The first occurs at $B\simeq 15.3$T, and the second at $B\simeq 21.5$T.

In discussing these results, the aim is to explore whether there is sufficient information in the spectra to make phase assignments upon varying the magnetic field. As discussed above, assigning UUD to the lowest trace is straightforward. In what follows, it will be useful to recognize the evolution of the Ba(2) components, even if we concentrate on Ba(1). Starting with 14.5T (bottom trace in Fig. \ref{fig:Sp_vsB_a_HFL12}), three peaks are observed, \textit{a}-\textit{c}. Peak \textit{a} comprises part of the Ba(1) spectrum, \textit{b} includes contributions from both Ba(1) and Ba(2), and \textit{c} is Ba(2) only. In the UUD phase in particular, one-third of Ba(1) nuclei see the ($\uparrow\uparrow$) configuration of ordered Co moments (peak \textit{a}) and two-thirds the ($\uparrow\downarrow$) configuration (peak \textit{b}). For the twice as abundant Ba(2), two-thirds contribute to peak \textit{c} and one-third overlap with Ba(1) on peak \textit{b}. Hence, the total relative intensity ratios expected for (\textit{a}:\textit{b}:\textit{c}) in the UUD phase are 1:4:4. On increasing the field beyond $\sim 15.5$T, the two inequivalent hyperfine fields of Ba(1) vary continuously from the fixed values seen in the UUD phase: the site with anti-aligned NN Co moments develops a stronger (more negative) hyperfine field, and the site with aligned NN moments experiences a weaker field. These results are fully consistent with the UIF phase, with spin configuration sketched on the right side of Fig. \ref{fig:Sp_vsB_a_HFL12} (also Fig. \ref{fig:states-3D}(b)). The field variations are associated with rotations of the aligned and anti-aligned moments away from the direction of the applied field, while also increasing the average hyperfine field. As the ground state in zero field is easy-plane, the rotations should be coplanar and orthogonal to the $\hat{c}$-axis. Although we don't go into detail here, the Ba(2)\textit{c} absorption peak split is also expected (see Fig.\ref{fig:SpectralIntegration}).

\begin{figure}[h]
\includegraphics[width=.35\textwidth]{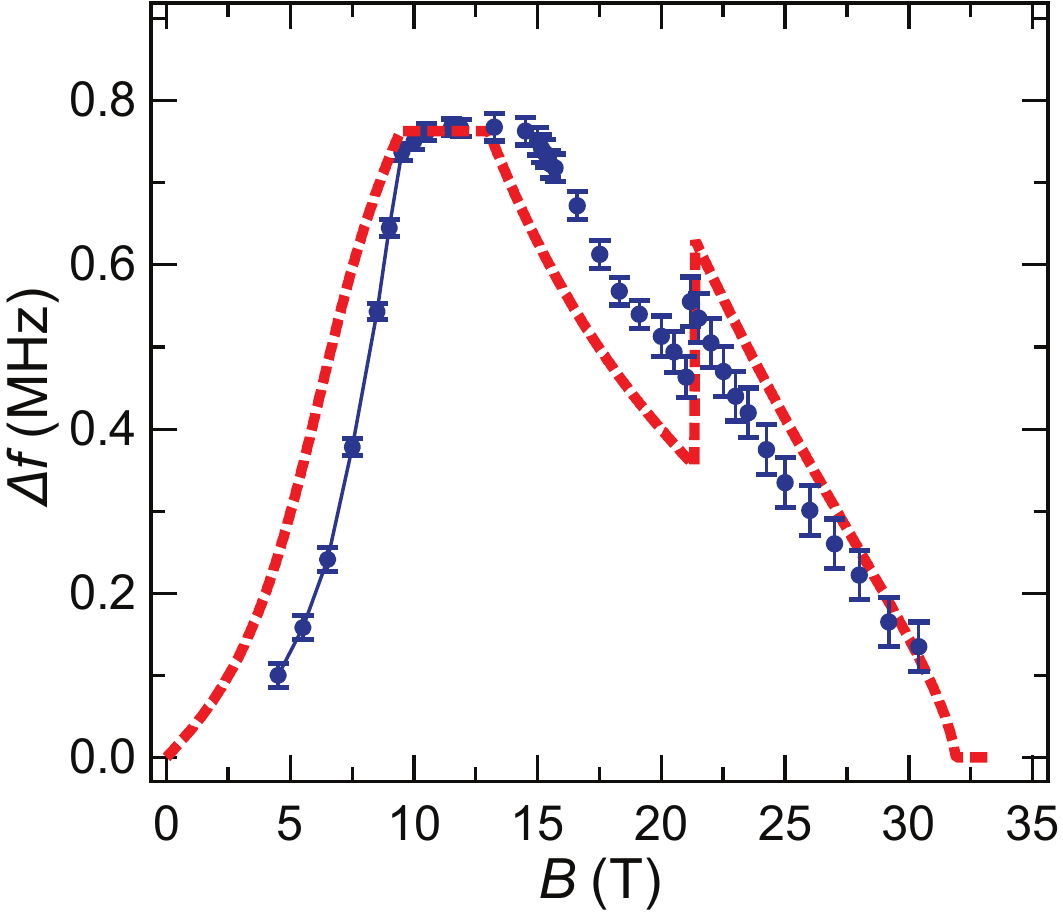}
\caption{Spectral frequency difference for Ba(1) site, plotted against applied magnetic field $\mathbf{B}\perp\hat{c}$. The red dashed line is the result from the model, and the data points are from the observed spectra. The plateau maximum corresponds to UUD. Spectral overlap with Ba(2) sites in spectral peak \textit{b} limits the accuracy (see Fig. \ref{fig:Sp_vsB_a_HFL12}); the error bars reflect the linewidth for the peak \textit{b}. }
\label{fig:Split_vsBpara}
\end{figure}

\begin{figure}[h]
\includegraphics[width=.38\textwidth]{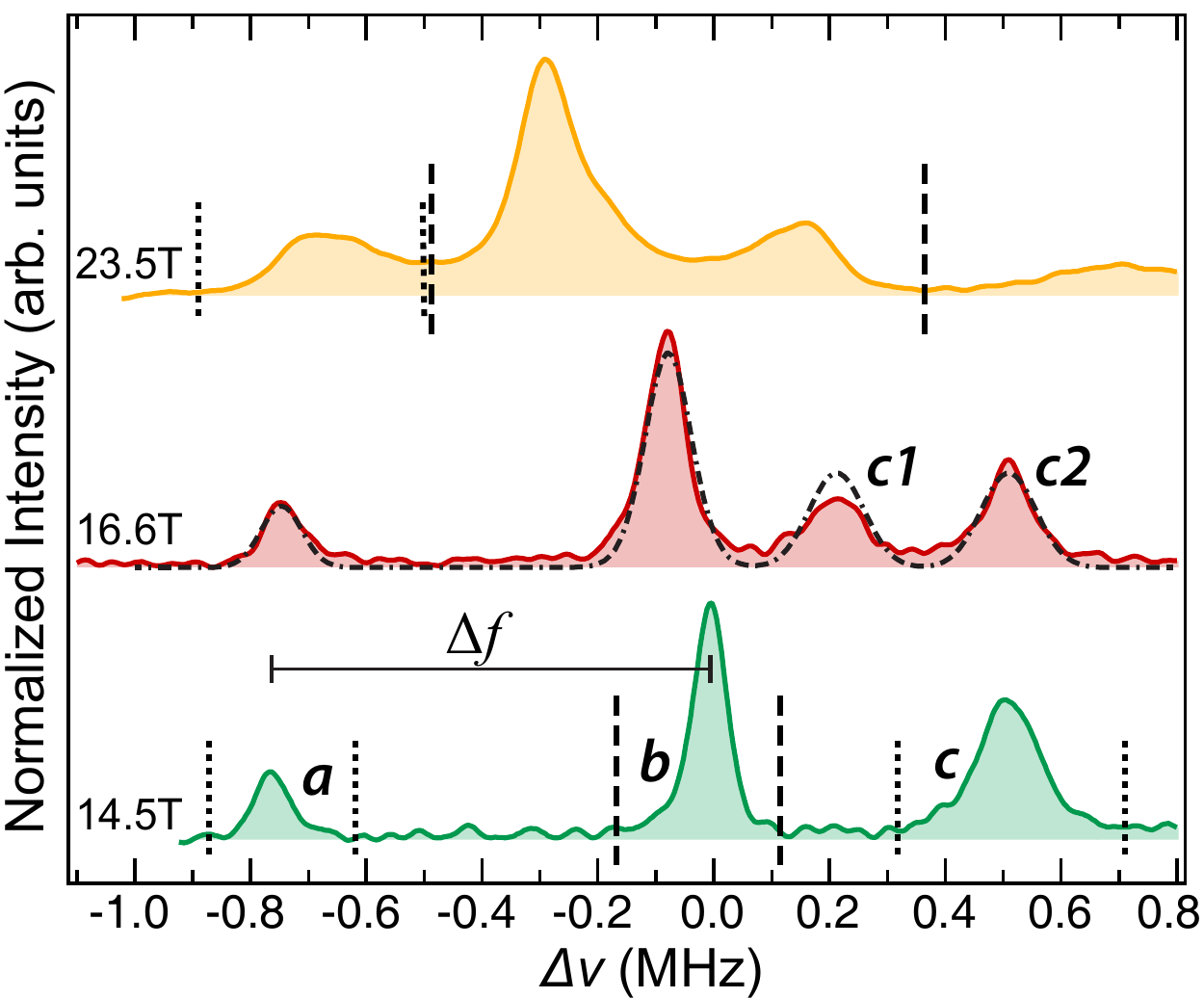} 
\caption{Sample $^{137}$Ba NMR spectra of the different accessed phases in Fig. \ref{fig:Sp_vsB_a_HFL12}. Dashed vertical lines denote the windowing used for evaluating the relative spectral intensity rearrangement for Ba(1) upon the UIF$\to$HF transition, listed in Table \ref{tab:IntegratedIntensities}. $\Delta f$ describes the Ba(1) site frequency difference, shown in Fig. \ref{fig:Split_vsBpara}. At 16.6T, the dashed line depicts the simulated spectrum of the Ba(1), Ba(2) sum, as expected in the UIF phase.}
\label{fig:SpectralIntegration}
\end{figure}

There is further mixing of Ba(1) and Ba(2) spectral components at higher fields, and this makes identification of the phase beyond the apparent transition at \mbox{$\sim$21.5T} ambiguous. Further complicating the issue, none of the features are fully resolved at the higher fields due to broadening of each of the components, combined with a trend toward spectral collapse beyond the saturation field. Nonetheless, we make a tentative identification of the transition as UIF$\to$HF. A quantitative comparison of the  expected field dependence of the Ba(1) spectral splitting ($\Delta f$) and the observed results is made in Fig \ref{fig:Split_vsBpara}. The red dashed line is generated from the model, whereas the data points are an attempt to follow the Ba(1) hyperfine fields. The $M_\text{sat}/3$ plateau phase has maximum splitting, which then decreases after entering UIF. The discontinuity occurs at the first order UIF$\to$HF transition. Even though the observed jump in $\Delta f$ is relatively small, the implications for the Ba(1) NMR spectrum are unmistakable: the relative intensities of the weak and strong hyperfine field sites are 2:1 in the UUD/UIF phases, whereas that intensity ratio reverses to 1:2 in the HF phase. Along with the intensity reversal must be a large change towards small shift for the site with weaker hyperfine field. Unfortunately, spectral overlap with Ba(2) sites, combined with field-dependent line-broadening limits accuracy. The error bars reflect the linewidth for the peak \textit{b}, which is used to estimate the uncertainty in the position of the relevant Ba(1) contribution.

The Ba(1) intensity reversal is revealed by isolating the relative intensity in the portion of the Ba(1) spectrum uncontaminated by Ba(2) contributions, i.e. peak \textit{a}. 
The relevant integration windows are shown in Fig. \ref{fig:SpectralIntegration}, where representative spectra of the three phases accessed in Fig. \ref{fig:Sp_vsB_a_HFL12} are depicted. In Table \ref{tab:IntegratedIntensities}, the integration is carried out for these spectra recorded at 14.5, 16.6, 23.5T, corresponding to UUD, UIF and (tentatively) HF phases. Spectra from other fields on both sides of the phase transitions display similar results. On passing to field strengths larger than 22T, the relative intensity of peak \ita\ increases sharply, consistent with the spin configuration rearrangement corresponding to a UIF$\to$HF transition.
\begin{table}[h]
\begin{tabular}{|p{.8cm}||p{.7cm}|p{.7cm}|p{.7cm}||p{.7cm}|p{.7cm}|p{.7cm}|}
\hline
  & \multicolumn{3}{|c|}{Observed intensity} & \multicolumn{3}{|c|}{Expected result}\\

\hline
$B$(T) & \textit{a} & \textit{b} & \textit{c} & \textit{a} & \textit{b} & \textit{c}\\
\hline\hline
14.5 & 1.1 & 3.9 & 4.1 & 1 & 4 & 4\\
\hline
16.6 & 1.1 & 4.1 & 1.8+2 & 1 & 4 & 2+2\\
\hline\hline
 & \textit{a} & \multicolumn{2}{|l|}{\textit{b+c}} & \ita & \multicolumn{2}{|l|}{\textit{b+c}}\\
\hline\hline
16.6 & 1.1 & \multicolumn{2}{|c|}{7.9} & 1 & \multicolumn{2}{|c|}{8}\\
\hline
23.5 & 1.7 & \multicolumn{2}{|c|}{7.2} & 2 & \multicolumn{2}{|c|}{7} \\
\hline
\end{tabular}
\caption{Integrated intensities, in reference to the spectra and labelling in Fig. \ref{fig:SpectralIntegration}. The designations \ita, \itb, \itc\ correspond to the integration ranges shown. The sum of the intensities is normalized to 9, corresponding to the combined contributions from Ba(1) sites (3) and Ba(2) sites (6). At 16.6T, the intensity of peak \textit{c} is written as the sum of the split components \textit{c}1,\textit{c}2.}\label{tab:IntegratedIntensities}
\end{table}

Still, some reasons for concern remain. Absent is a clear signature of a Ba(1) spectral peak near zero frequency upon increasing the field through the UIF$\to$HF transition. Also, the broadening of all of the spectral features add to the challenge of a definitive identification at the higher fields, which also bear some resemblance to what would result from incommensurate structures. Further clouding the issue is the absence of any substantial first order character in the specific heat data \cite{FortuneAPS}, as well as the apparently continuous magnetization results~\cite{Susuki:2013}. Both are consistent with a second order transition. Unfortunately, we cannot distinguish between the two possibilities with the existing spin lattice relaxation results.

\subsection{%
  NMR near ${\mathbf B}\parallel\hat{c}$
  \label{subsec:H||c,large-H}
}
\subsubsection{Low-field Measurements ($B<12\rm{T}$) }
\begin{figure}[h]
\includegraphics[width=3.3in]{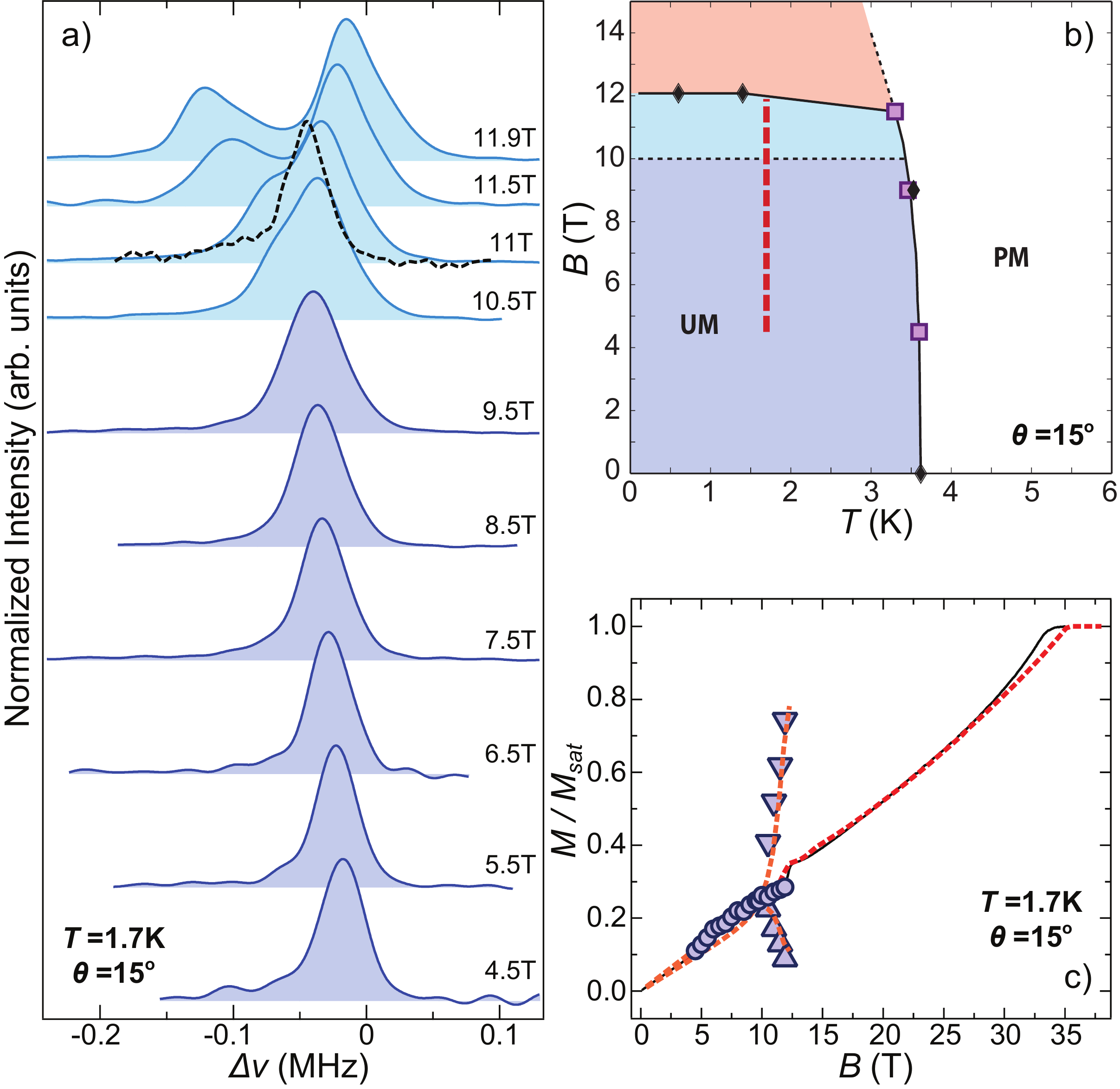}
\caption{(a) Field evolution of the $^{137}$Ba(1) hyperfine frequency shifts (see text) for $\theta=15^{\circ}$ at $T=1.7$K. The black dashed line shows the spectrum with absence of splitting for $\theta=0^{\circ}$ and $B$=11T. (b) Experimental phase diagram for ${\mathbf B}\parallel\hat{c}$, from magnetization results (black diamonds).\cite{HaidongM} Squares reflect $T_N$ as measured by $T_1^{-1}(T)$ (Fig.~\ref{fig:BciTone}). The heavy dashed line corresponds to the field range and temperature covered in (a). The different color for fields $B>10$T is not for a distinct phase, but corresponds to the onset of the double peak structure for the Ba(1) spectrum seen in (a). (c) Magnetization vs.~magnetic field.  The solid, dashed lines are from magnetization results, and the model of Eq.~\eqref{eq:MF}, respectively. The data points are derived from the NMR spectra: circles are the first moment of the full spectrum, properly normalized, and the triangles are associated to the hyperfine shifts of the two Ba(1) local environments.}
\label{fig:BcSpectra}
\end{figure}

In turning to the results from fields applied out-of-plane, the phase diagram used for comparison, derived from magnetization measurements with ${\mathbf B}\parallel\hat{c}$, is shown in Fig.~\ref{fig:BcSpectra}b. In this set of measurements, the field alignment was approximately $\theta=15^\circ$ from the $\hat{c}$-axis. While we presume that many aspects of the $B$-$T$ phase diagram obtained from ${\mathbf B}\parallel\hat{c}$ still apply, there were nevertheless additional features observed in the Ba(1) spectra due to the deviation from $\hat{c}$-axis that we describe below.

The $^{137}$Ba(1) central transition spectra are shown in Fig.~\ref{fig:BcSpectra}a for $\theta=15^{\circ}$ at $T=1.7$K. On increasing the applied field from 4.5T, a single line is observed until the field reaches $\sim$10T. Further increase in the magnetic field leads to line broadening, and subsequently a double-peak structure is evident. The single line spectrum is distinctive of the umbrella phase (Fig. \ref{fig:states-3D}a(a)), with the angle of the umbrella closing for increasing field towards its direction. As illustrated in  Fig.~\ref{fig:BcSpectra}b, a line splitting occurs at the same field independent of temperature. This is the result of the spin structure's distortion from umbrella when moving off $\hat{c}$-axis. In particular, the specific non-coplanar configuration realized in this case (see Fig.~\ref{fig:LFC}) leads to two separate local field environments for the Ba(1) nuclear sites, and hence broadening of the NMR line.  In Fig.~\ref{fig:BcSpectra}c, the evolution of the hyperfine fields is plotted and compared to that produced by the model for a similarly aligned magnetic field. Note that the development of inequivalent Ba(1) sites thus occurs only on approaching the actual phase transition, both in the experiment and the model. That is to say, the distortions are nonlinear.

In presenting the results for $\theta=15^{\circ}$, we have suggested that so far as the magnetic field vs.~temperature phase diagram is concerned, not much is changed from $\theta=0$, save for the distortion which becomes more severe as the first order transition at $B\sim12$T is approached. The model also suggests a possibility of second-order transition when the field orientation deviates from the $\hat{c}$-axis beyond a critical value (see Fig. \ref{fig:phase-diagram}). Due to field limitations, this was not explored.

\begin{figure}[h]
\includegraphics[width=2.3in]{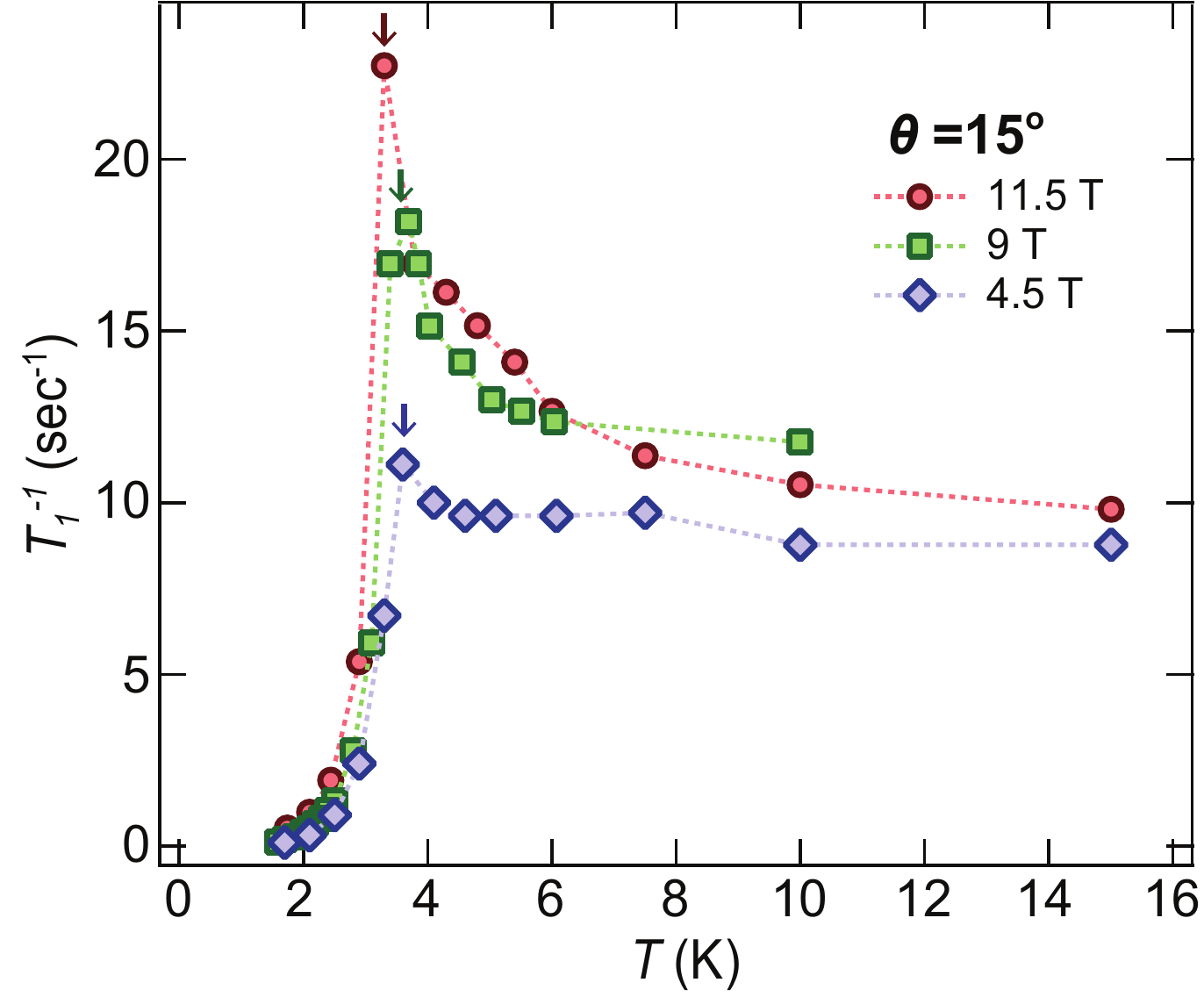}
\caption{Spin-lattice relaxation rate $T_1^{-1}(T)$ at various applied field values and orientation $\theta=15^{\circ}$. The arrows indicate the transition temperature $T_N$.}
\label{fig:BciTone}
\end{figure}

In Fig.~\ref{fig:BcSpectra}b, the solid square data points were derived from measurements of the spin-lattice relaxation rate $T_1^{-1}(T)$. The temperature dependence, covering the transition, is shown in Fig.~\ref{fig:BciTone}. We take the sharp drop upon cooling as the indicator for a first order transition at $T_N(B)$ at each of the measured fields, and otherwise note that very little increase in rate is observed as $T_N$ is approached from higher temperatures. Below the ordering temperature, the drop in relaxation rate occurs quite fast. At lower temperatures, the variation is indistinguishable from an activated form with $\Delta\sim 11$K, though this seems too large relative to $J$ to be induced by the deviation of $\mathbf B$ from the $\hat{c}$-axis. With anisotropic contributions to the hyperfine coupling, two-magnon processes are possible and lead to $1/T_1\sim T^d$, with $d$ the dimensionality. We acknowledge that some caution is warranted here, since the variation of the temperature relative to the ordering is limited.

\subsubsection{High-field Measurements ($B>14\rm{T}$)}
\begin{figure}[h]
\includegraphics[width=2.6in]{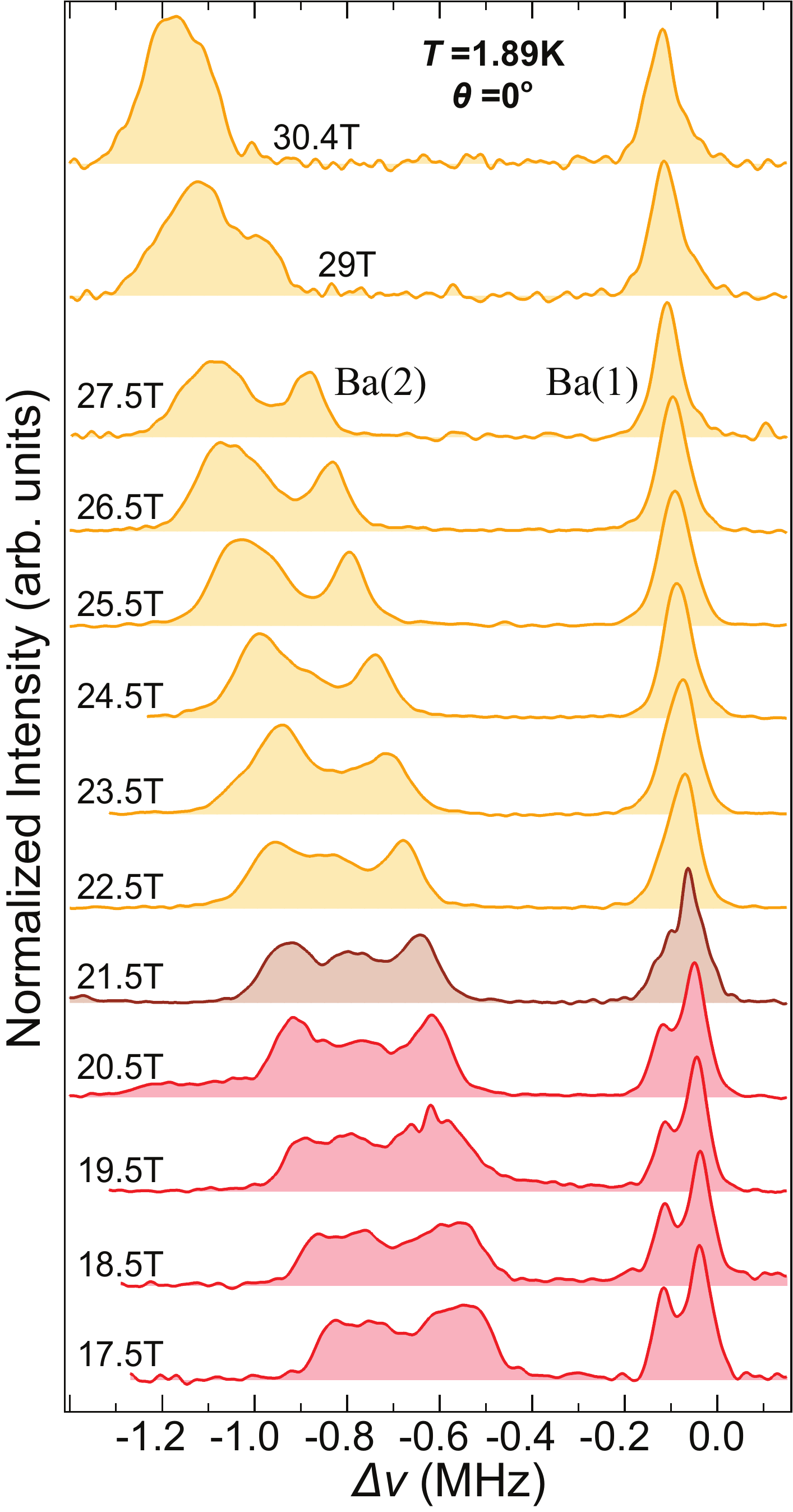}
\caption{Spectra vs Field for $\mathbf{B}\parallel\hat{c}$. The Ba(1) and Ba(2) contributions are well-separated. For field strengths less than \mbox{$\sim$22T}, the Ba(1) line exhibits a doubled-peak structure, consistent with the prediction for the LIF phase. This splitting appears to collapse at \mbox{$\sim$22T}, and, simultaneously, an asymmetry develops in the Ba(2) contribution.}
\label{fig:Sp_vsB_c_HFL12}
\end{figure}

The NMR spectra for the central transition of Ba(1) and Ba(2) in larger fields aligned with the $\hat{c}$-axis are shown in Fig. \ref{fig:Sp_vsB_c_HFL12}. For Ba(1), observed are two broad lines at lower field, collapsing to a single line, also fairly broad, at higher fields. A phase transition is seen at a lower field in magnetization studies, at $\sim$12T; this is not covered by the dataset.

\begin{figure}[h]
\includegraphics[width=2.55in]{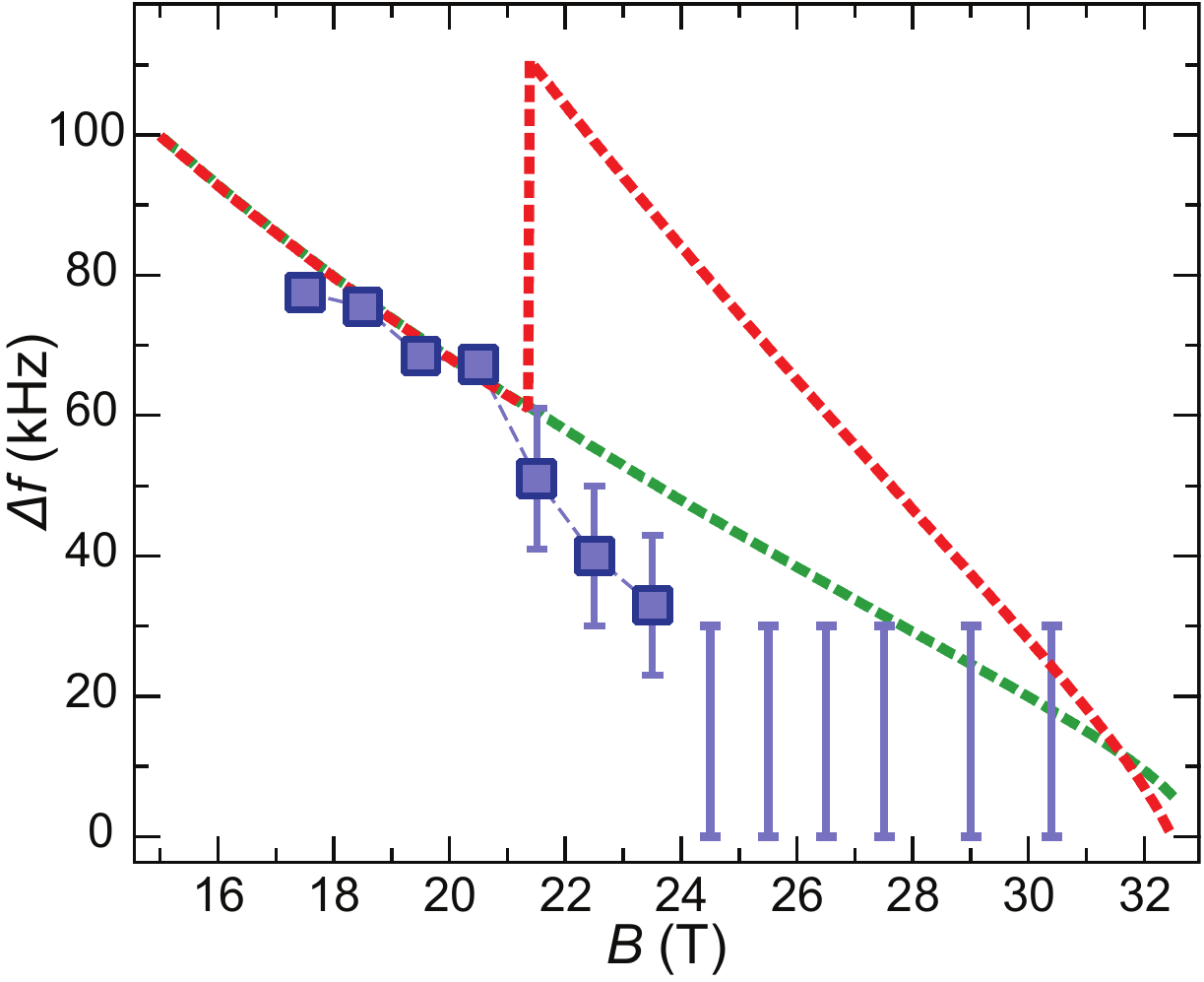}
\caption{
  Ba(1) spectral line splitting vs applied field for $\mathbf{B}\parallel\hat{c}$, extracted from the data shown in Fig. \ref{fig:Sp_vsB_c_HFL12}. The red dashed line is what is expected assuming the LIF$\to$HF transition occurs at 22T, whereas the green dashed line is an extrapolation of the Ba(1) splitting if the LIF phase remains stable and no transition occurs. Heavy blue lines above 24T denote an estimate of the minimum splitting that can be resolved due to the spectral linewidth. Thus, beyond 24T, only a single absorption peak is observable within our experimental sensitivity.
}
\label{fig:Split_vsBparc}
\end{figure}

The model calculation predicts 2 phase transitions upon increasing the magnetic field, before reaching saturation at $\sim$33T. As discussed earlier in the context of lower fields, the first is a transition from the umbrella to the coplanar LIF phase at \mbox{$B\sim$12T}, which is expected first order provided that the magnetic field is aligned sufficiently close to the $\hat{c}$-axis. Consistent with the data covering intermediate fields 14-22T, the Ba(1) lineshape is expected to reveal two local environments of distinctive hyperfine field, with intensity ratio of 1:2. Here, the observed resonance frequency difference between the two relevant peaks, \mbox{$\Delta f=$ $\gamma \Delta B_{hf}$}, is considerably smaller than in the case of similar spin configuration for $\mathbf{B}\perp\hat{c}$, due to the smaller hyperfine coupling constant \cite{Appendix2}. This somewhat limits our ability to resolve the distinct Ba(1) contributions when their respective hyperfine field difference becomes smaller. A second transition  is anticipated at \mbox{$B\sim$22T} to the HF phase, and again two local environments are expected for Ba(1). However, only a single broad Ba(1) line is observed beyond 22T, with its linewidth being much narrower than the expected splitting. This is illustrated in Fig. \ref{fig:Split_vsBparc}, where the red dashed line depicts the model prediction for a transition from LIF$\to$HF at 22T. Instead of the expected large increase, to within the resolution limits, the Ba(1) peak separation is observed to collapse. The filled blue squares show the separation, obtained from a fit to the spectrum consisting of two components of Lorentzian lineshape. We take the collapse as evidence for a phase transition, but the ground state for $B\gtrsim$22T is inconsistent with the predicted HF phase (see Fig. \ref{fig:phase-diagram}, Fig. \ref{fig:states-3D}(c)).

In principle, taking into account the spectra from Ba(1) and Ba(2) sites could help to reduce the possible phases for this regime. Specifically, assuming that the Ba(1) spectrum above \mbox{$\sim$22T} indeed represents a single hyperfine field environment, the only consistent with the data state among the candidate ones shown in Fig. \ref{fig:states-3D} is the ``umbrella'', which results in the same local field for all Ba(1) nuclei. However, for such a state, the Ba(2) spectrum is expected to be symmetric for $\mathbf{B}\parallel\hat{c}$. Since the observed Ba(2) lineshape is asymmetric, we leave this issue unresolved for the moment. 

Finally, we note that the spectral lineshape of the Ba(2) contribution seems to suddenly change at 29T, even though the available data in this field region is limited. This could  be simply reflecting the tendency of the sublattice moments to fully align with the applied field approaching the saturation value, but it could also indicate an additional phase transition at high fields and before reaching saturation. Further measurements are required to address this possibility. 

\section{Conclusion}
In concluding, we have carried out \ba\ NMR measurements on the easy-plane triangular lattice antiferromagnet \baco\ to fields ranging to 30T, which is just short of the saturation field at 33T. For fields aligned in-plane, 4 phases are accessed in the process of approaching the saturation magnetization, which leave their identifiers in the Ba NMR spectra. In addition, the phase transitions are distinguished by the field dependence of the spin-lattice relaxation rate. Three phases are accessed for fields aligned with the $\hat{c}$-axis. The experimental results are found to be in excellent agreement with the predictions of a semiclassical treatment developed for computing the quantum phase diagram of \baco, which incorporates the effect of quantum fluctuations via the generation of effective coupling constants for the classical spins. A clear deviation between the predictions and the experimental outcome is observed for fields exceeding 22T, aligned along $\hat{c}$-axis.

\section{Acknowledgements}
The authors are grateful to Nat Fortune, Oleg Starykh, Yasu Takano, and Hidekazu Tanaka for fruitful discussions, and to Nat Fortune and Yasu Takano for sharing their results prior to publication. The research reported here was supported by the National Science Foundation under grant no. DMR-1105531 (UCLA). Work at LANL was performed under the auspices of the U.S.\ DOE  through the Laboratory Directed Research and Development program. G.K. acknowledges support from the LANL Seaborg Institute, and Y.K. acknowledges support from the RIKEN iTHES Project.

\section{Appendix}
\begin{figure}[h]
\includegraphics[width=3.2in]{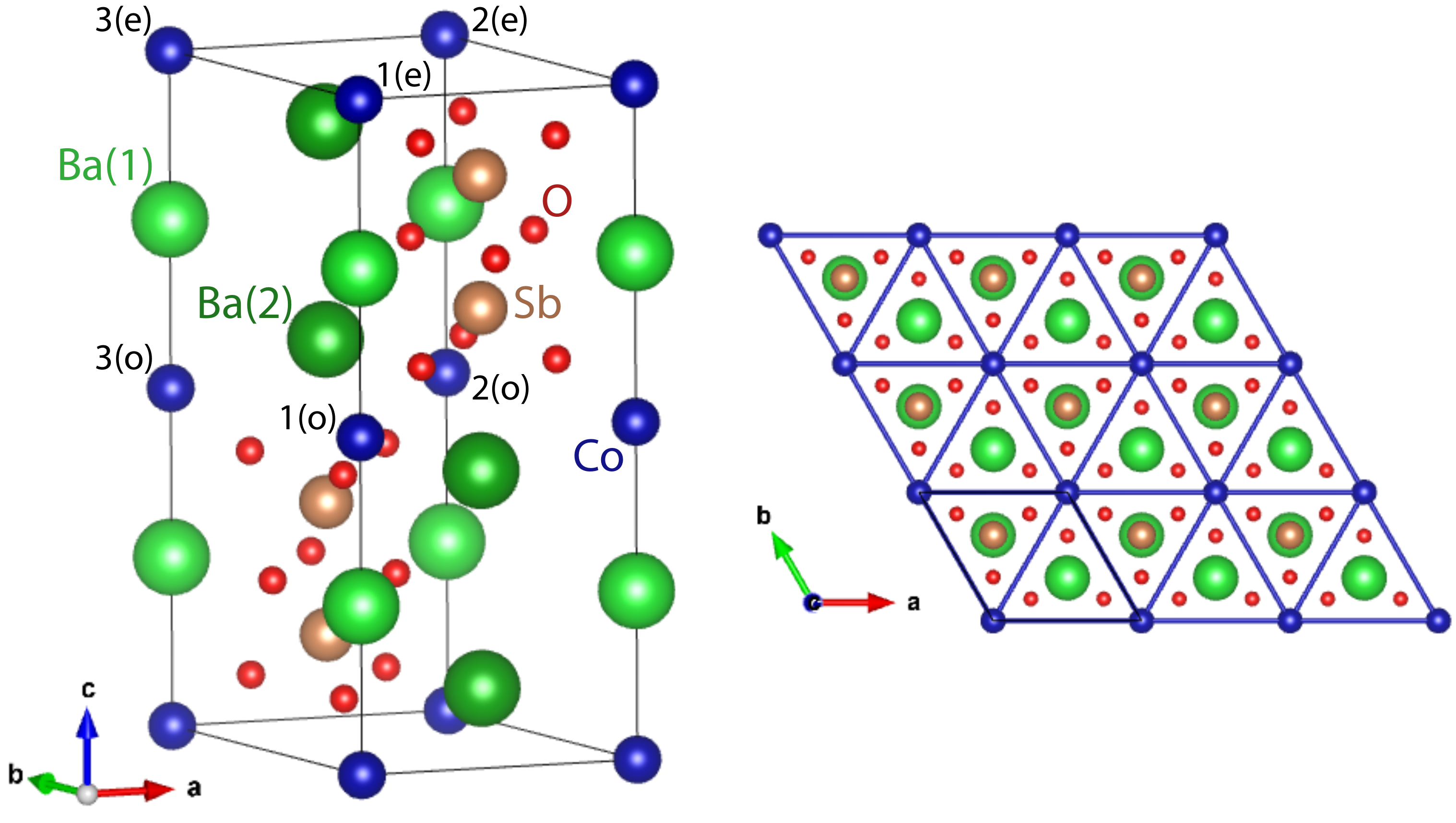}
\caption{Chemical unit cell of \baco . Co$^{2+}$ ions (blue) occupy the edges of triangular lattices (1-3) on different layers (odd-even) in the $ab$-plane. There are two inequivalent Ba sites, Ba(1) (light green) and Ba(2) (dark green).}
 \label{fig:xtal}
\end{figure}
Figure~\ref{fig:xtal} depicts the crystal structure of \baco. It represents a highly symmetric hexagonal structure, space group $P6_{3}/mmc$, with lattice constants $a=b=5.8562$\AA\ and $c=14.4561$\AA. Layers of regular magnetic triangular lattices are formed parallel to the $ab$-plane by the \mbox{Co$^{2+}$} ions (blue), which carry effective spin $S=1/2$. The different Co sublattices are denoted by 1-3 on even ($e$), odd ($o$) layers. There are two inequivalent, uniaxially symmetric Ba sites in the unit cell, referred to as Ba(1) (light green) and Ba(2) (dark green).

\begin{figure}[h]
\includegraphics[width=3.3in]{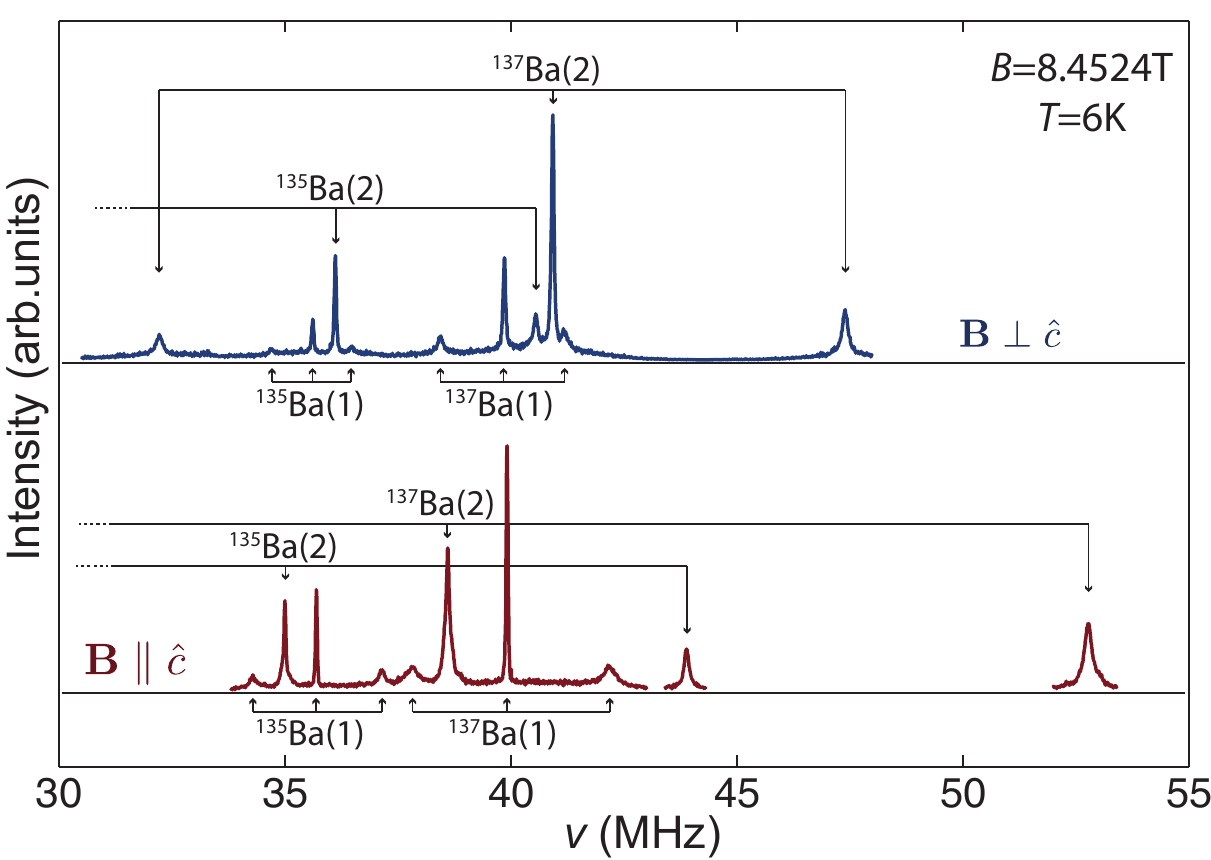}
\caption{\baco\ NMR spectrum for $\mathbf{B}=8.4524$T $\parallel,\perp\hat{c}$ at $T=6$K. For each Ba site/isotope, the spectrum is composed of three quadrupole transition lines (Eq. \ref{eq:NMR}), the central $\left<1/2\leftrightarrow -1/2\right>$ and the two satellites $\left<\pm 3/2\leftrightarrow \pm 1/2\right>$. The difference in frequency between the two satellites is equal to $\nu_Q$ ($2\nu_Q$) for $\mathbf{B}\perp\hat{c}$ ($\mathbf{B}\parallel\hat{c}$).}
 \label{fig:spectra}
\end{figure}
The full NMR hamiltonian for a nucleus of spin $I\neq1/2$ and gyromagnetic ratio $\gamma$, sitting on an axially symmetric site, is given by
\begin{equation}
\mathcal{H}=-\gamma \hbar \mathbf{I}\cdot \left(\mathbb{I}+\mathbb{K}\right)\cdot\mathbf{B}+\frac{h \nu_Q}{6} \left[ 3 I_z^2 -  I(I+1) \right],
\label{eq:NMR}
\end{equation}
where $\mathbf{B}$ is the applied magnetic field, $\mathbb{K}$ is the NMR shift tensor, and $\nu_Q$ is the nuclear quadrupole frequency. For $I=3/2$, three NMR lines occur, corresponding to the nuclear transitions $\left<m_I\leftrightarrow m_I-1\right>$. There are two NMR active Ba isotopes, \baF\ with $^{135}\gamma/2\pi=4.2295$MHz/T and \baS\ with $^{137}\gamma/2\pi=4.73158$MHz/T, both with $I=3/2$. Their respective abundances are 6\% and 11\%. Thus, in \baco\ where there are two crystallographically inequivalent Ba sites, four sets of three lines are expected in the NMR spectrum. Figure~\ref{fig:spectra} shows the NMR spectrum in the paramagnetic state for $\mathbf{B}\parallel,\perp\hat{c}$, with the various lines labeled accordingly. The corresponding values of $\nu_Q$ are listed in table \ref{tbl:NMRpar}.
\begin{table}[!hbp]
\begin{tabular}{|c|c|c|c|}
\hline
Nuclear site  & $\nu_Q$(MHz) & $A_{\parallel c} (G/\mu_B)$ & $A_{\perp c} (G/\mu_B)$\\
\hline
$^{137}$Ba(1) & 2.72 & -156 & -612 \\
\hline
$^{137}$Ba(2)& 15.4 & -1313 & -752 \\
\hline
$^{135}$Ba(1) & 1.75 & -156 & -612 \\
\hline
$^{135}$Ba(2) & 9.8 & -1313 & -752 \\
\hline
\end{tabular}
\caption{NMR parameters for the Ba nuclear sites/isotopes.}
\label{tbl:NMRpar}
\end{table}
\begin{figure}[h]
\includegraphics[width=2.4in]{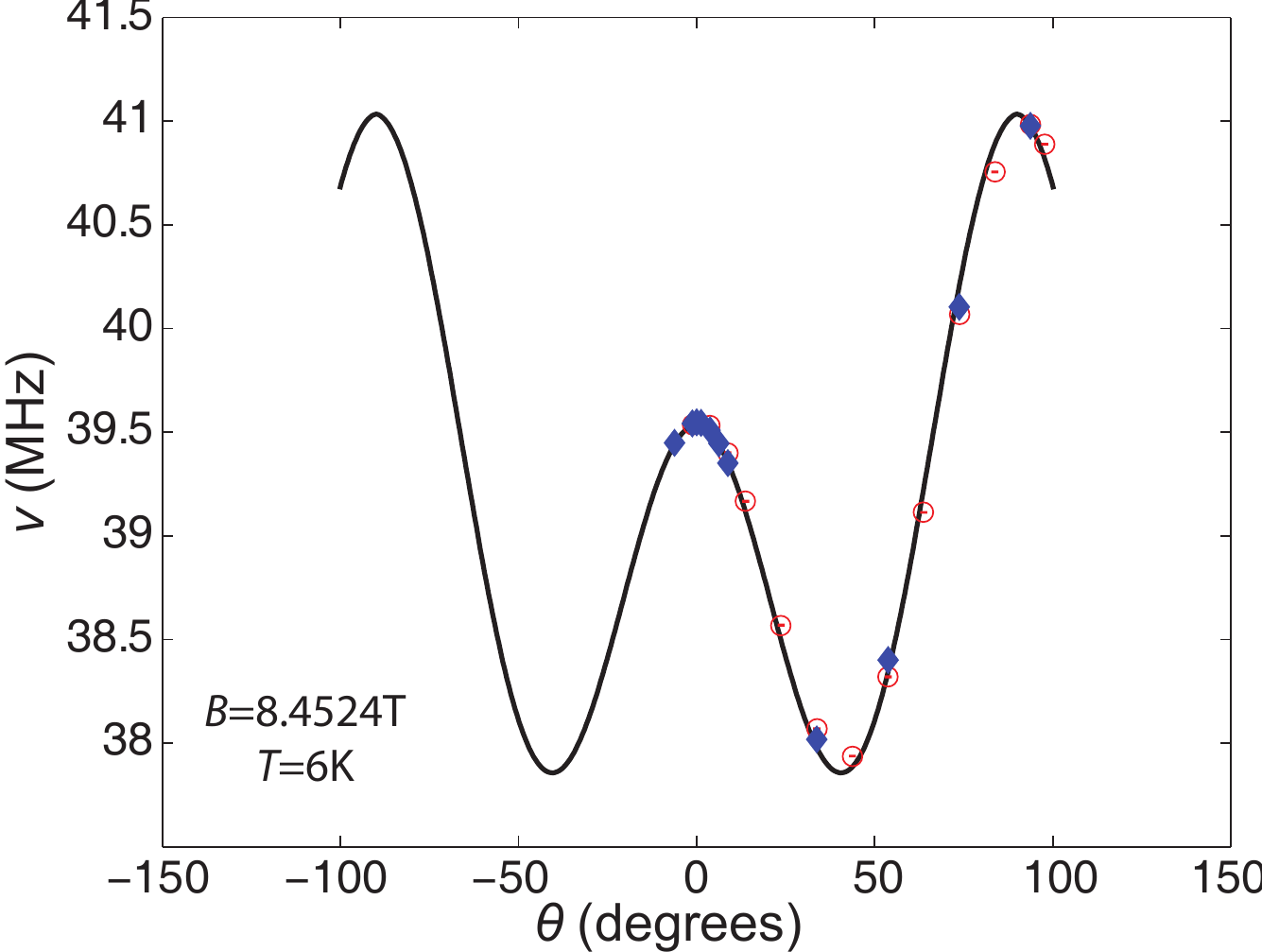}
\caption{NMR frequency of the $^{137}$Ba(2) central transition {\it vs.}~$\theta$ for $B=8.4524$T and $T=6$K.}
 \label{fig:rotation}
\end{figure}
The crystal orientation was verified by mapping the resonance frequency of the $^{137}$Ba(2) nuclear central transition for a rotating applied field in the $ac$-plane. The results are plotted in Fig.~\ref{fig:rotation} as a function of the angle $\theta$ from the $\hat{c}$-axis. Red (blue) data points correspond to experimental runs for rotation with increasing (decreasing) $\theta$. The solid line shows the relevant expected frequency value as of Eq. \ref{eq:NMR}, which up to second order in $\nu_Q$ is given by $\nu=\gamma B\left(1+K_asin\theta+K_c cos\theta\right)+\frac{3\nu_Q^2}{16\gamma B}\left(1-cos^2\theta \right)\left(1-9cos^2\theta \right)$. The shift components $K_a$, $K_c$ are determined by the local maxima positions in Fig.~\ref{fig:rotation}.

\begin{figure}[h]
\includegraphics[width=3.2in]{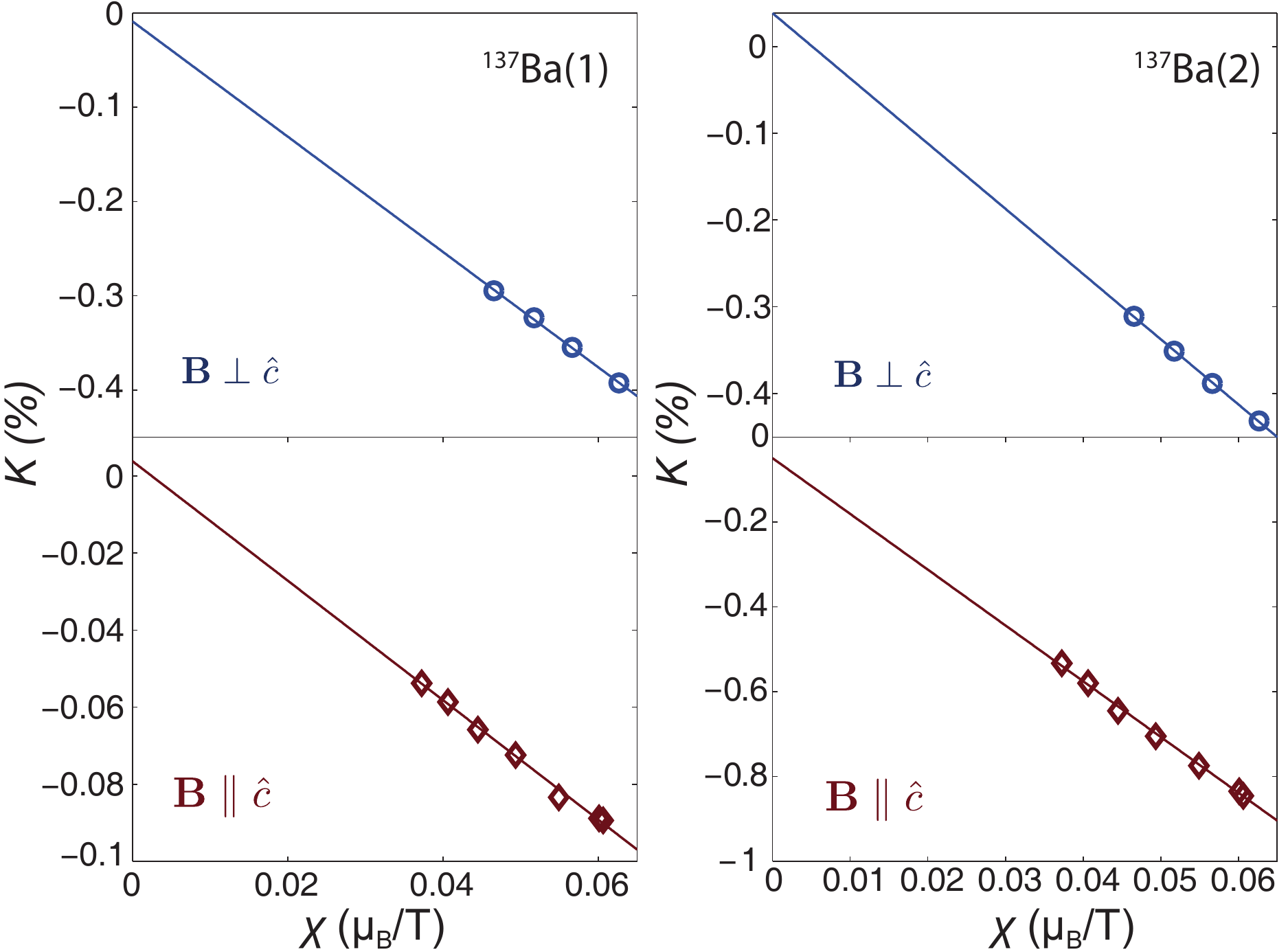}
\caption{$K-\chi$ plots for both Ba sites at $\mathbf{B}=8.4524$T$\parallel,\perp\hat{c}$.}
 \label{fig:Kchi}
\end{figure}

In the paramagnetic state, the NMR shift $K_{\alpha}$ is given by $K_{\alpha}(T)=K_{\alpha}^{0}+A_{\alpha}\chi_{\alpha}(T)$, where $\alpha$ denotes the field orientation, $K_{\alpha}^{0}$ is the temperature independent orbital contribution, $A_{\alpha}$ is the hyperfine coupling constant, and $\chi_{\alpha}$ the susceptibility. Thus, $A_{\alpha}$ can be deduced from the slope of a $K-\chi$ plot, for which the implicit parameter is temperature, and $K_{\alpha}^{0}$ from the y-intercept. The results are shown in Fig.~\ref{fig:Kchi} for the different Ba isotopes and nuclear sites, and the relevant values for $A_{\alpha}$ are listed in table \ref{tbl:NMRpar}.
\newpage

\bibliographystyle{apsrev4-1}

\begin{thebibliography}{35}
\bibitem [{App({\natexlab{a}})}]{Appendix1}%
  \BibitemOpen
  \href@noop {} {}\bibinfo {note} {The crystal structure, with emphasis on the relative positions of the Co ions and the Ba nuclear spin sites, is provided in the Appendix.}\BibitemShut {Stop}%
\end{thebibliography}

\end{document}